\documentclass[fleqn,usenatbib]{mnras}

\usepackage{newtxtext,newtxmath}
\usepackage[T1]{fontenc}

\usepackage{graphicx,xspace,lscape}
\usepackage[export]{adjustbox}

\newcommand{\fig}[2][]{Figure#1~\ref{fig:#2}}

\newcommand{\sect}[2][]{Section#1~\ref{sec:#2}}

\newcommand{\U}[1]{\ensuremath{\mathrm{~#1}}}     
\newcommand{\yr}{\U{yr}}
\newcommand{\Myr}{\U{Myr}}          
\newcommand{\Gyr}{\U{Gyr}}          
\newcommand{\pc}{\U{pc}}
\newcommand{\kpc}{\U{kpc}}
          
\newcommand{\Msun}{\U{M}_{\odot}\xspace}   
\newcommand{\Msunyr}{\Msun\yr^{-1}} \newcommand{\msunyr}{\Msunyr}
\newcommand{\Zsun}{\U{Z}_{\odot}}   
\newcommand{\cc}{\U{cm^{-3}}}
\newcommand{\K}{\U{K}}
    
\newcommand{\kms}{\U{km\ s^{-1}}}
\newcommand{\erg}{\U{erg}}

\newcommand{\mstar}{\ensuremath{m_\mathrm{\star}}}

\newcommand{\stir}{{\tt stir}\xspace}
\newcommand{\kick}{{\tt kick}\xspace}
\newcommand{\nofb}{{\tt no\,feedback}\xspace}
\newcommand{\launching}{{\it launching}\xspace}
\newcommand{\innercgm}{{\it inner-outer halo}\xspace}
\newcommand{\inferno}{{\small INFERNO}\xspace}
\newcommand{\ramses}{{\sc Ramses}\xspace}
\newcommand{\ramsesrt}{{\sc Ramses-rt}\xspace}
\newcommand{\arepo}{{\sc Arepo}\xspace}
\newcommand{\cloudy}{{\sc Cloudy}\xspace}

\newcommand{\nugrid}{{\small NuGrid}\xspace}

\newcommand{\fire}{{\small FIRE2}\xspace}

\newcommand{\toc}{\par\begin{center}\vspace{0.5mm}\begingroup\small\let\cleardoublepage\relax\let\clearpage\relax\mytoc\endgroup\vspace{0.5mm}\end{center}\par} 
 \makeatletter\newcommand\mytoc{\@starttoc{toc}}\makeatother 
\long\def\symbolfootnote[#1]#2{\begingroup%
\def\thefootnote{\fnsymbol{footnote}}\footnote[#1]{#2}\endgroup} 

\newcommand{\lund}{Department of Astronomy and Theoretical Physics, Lund Observatory, Box 43, SE-221 00 Lund, Sweden}
\newcommand{\princeton}{Department of Astrophysical Sciences, Princeton University, Princeton, NJ 08544, USA}
\newcommand{\amnh}{Department of Astrophysics, American Museum of Natural History, 200 Central Park West, New York, NY 10024, USA}

\title[INFERNO]{INFERNO: Galactic winds in dwarf galaxies with star-by-star simulations including runaway stars}

\author[E. P. Andersson et al.]{
Eric P. Andersson$^{1,2}$\thanks{E-mail: eandersson@amnh.org}, 
Oscar Agertz$^{2}$, 
Florent Renaud$^{2}$ and
Romain Teyssier$^{3}$
\\
$^{1}$\amnh \\
$^{2}$\lund \\
$^{3}$\princeton
}

\date{Accepted XXX. Received YYY; in original form ZZZ}

\pubyear{2023}

\begin{document}
\label{firstpage}
\pagerange{\pageref{firstpage}--\pageref{lastpage}}
\maketitle

\begin{abstract}
The formation and evolution of galaxies have proved sensitive to the inclusion of stellar feedback, which is therefore crucial to any successful galaxy model. We present \inferno, a new model for hydrodynamic simulations of galaxies, which incorporates resolved stellar objects with star-by-star calculations of when and \emph{where} the injection of enriched material, momentum, and energy takes place. \inferno treats early stellar kinematics to include phenomena such as walkaway and runaway stars. We employ this innovative model on simulations of a dwarf galaxy and demonstrate that our physically motivated stellar feedback model can drive vigorous galactic winds. This is quantified by mass and metal loading factors in the range of $10-100$, and an energy loading factor close to unity. Outflows are established close to the disc, are highly multi-phase, spanning almost $8$ orders of magnitude in temperature, and with a clear dichotomy between mass ejected in cold, slow-moving ($T\lesssim5\times10^4\K$, $v<100\kms$) gas and energy ejected in hot, fast-moving ($T>10^6\K$, $v>100\kms$) gas. In contrast to massive disc galaxies, we find a surprisingly weak impact of the early stellar kinematics, with runaway stars having little to no effect on our results, despite exploding in diffuse gas outside the dense star-forming gas, as well as outside the galactic disc entirely. We demonstrate that this weak impact in dwarf galaxies stems from a combination of strong feedback and a porous interstellar medium, which obscure any unique signatures that runaway stars provide. 
\end{abstract}




\section{Introduction}\label{sec:intro}

Galactic evolution is governed by a manifold of connected processes over a vast range of physical scales. An important aspect of this evolution and an example of this scale-coupling is galactic scale winds driven by feedback processes in the interstellar medium (ISM). This generates a baryon cycle \citep[for a review, see][]{Veilleux+2005,Zhang2018}. Understanding the injection of energy and momentum on parsec scales \citep{McKee&Ostriker1977,Katz1992,Kim&Ostriker2015}, how this translates into outflows \citep{Schroetter+2016,Chisholm+2017,Fielding+2017}, and how the ejected material evolves after leaving the galaxy \citep{Tumlinson+2017,Fielding2020} are fundamental questions for galaxy theory. To tackle these questions, semi-analytical models \citep[e.g.,][]{Baugh2006,Benson2010}, large-scale cosmological simulations \citep[e.g., ][]{EAGLE2015,Illustris2014,IllustrisTNG2019}, and simulations of the ISM \citep[e.g.,][]{Walch+2015,Kim+2020a,Kim+2020b} have been employed. Progress made towards answering these questions can be attributed both to advances in numerical methods and modeling, as well as an improved understanding of the physics involved \citep[see][for a review]{SomervilleDave2015}. The complex physics of gaseous material innate to these problems make hydrodynamic simulations combined with sub-grid models for the relevant unresolved physics highly suitable for this task \citep[see, e.g.,][]{Wheeler+2019,Agertz+2020,Smith+2021}. The recent progress made with these kinds of models has in part been facilitated by galaxy-scale simulations reaching a higher resolution, thereby better resolving processes within the ISM (and consequently the star-forming clouds), while capturing the global dynamics of evolving galaxies.

Today, galaxy simulations routinely reach a parsec-scale resolution, with star-particles representing individual stars \citep[see, e.g.,][]{Hu+2016,Emerick+2018,Andersson+2020,Andersson+2021,Hirai+2021,Gutcke+2021,Hislop+2022}, and in fact, \emph{should} be done in this way to avoid the many restrictions (e.g., location of individual stars) imposed by the traditional approach\footnote{To relieve the computational cost of tracking the vast number of stars hosted by galaxies, stars are typically modeled as single stellar populations which are tracked by a single particle. This approach becomes less sensible when the mass of the star particles is smaller than that of individual stars, which is often the case in highly resolved simulations.}. Star-by-star models allow for a detailed account of when and \emph{where} individual stars inject momentum, energy, and enriched material. The locality of supernovae (SNe) has already been shown to affect the efficiency of stellar feedback \citep[e.g.,][]{Walch+2015,Girichidis+2016,Gatto+2017}, in turn altering the properties of massive galaxies \citep[e.g.,][]{Ceverino&Klypin2009,Kimm&Cen2014,Andersson+2020} and dwarf galaxies \citep[e.g.,][]{Gutcke+2022,Steinwandel+2022}. This indicates that star-by-star models are necessary to fully study cloud evolution, star formation, stellar feedback, chemical mixing, and thus galaxy evolution as a whole.

To this end, we present the INdividual stars with Feedback, Enrichment, and Realistic Natal mOtions (\inferno) model, a new versatile star-by-star model implemented in the $N$-body+hydrodynamics code \ramses\citep{Teyssier2002}. The \inferno model is a step towards a complete account of the physics that drives galaxy formation and evolution. In its current state, \inferno accounts for feedback processes from giant branch stars, the rapidly evolving O and B type stars, core-collapse supernovae (CCSNe), and type Ia supernovae (SNeIa). Furthermore, \inferno treats the early collisional dynamics in natal star clusters, which is one origin of walkaway and runaway stars \citep{Poveda+1967,OhKroupa2016}.  

How massive runaway stars affect galaxy evolution is still a debated question \citep[for a review, see][]{NaabOstriker2017}. These types of fast-moving stars are key examples of processes that relocate SNe. As mentioned before, this affects the stellar feedback and as a result outflows. This, often called \emph{random versus peak driving}, has been explained by the interplay between clustered star formation \citep[and consequently clustered feedback of short-lived stars, see e.g.,][]{MacLow&McCray1988,Nath&Shchekinov2013,Sharma2014,Keller+2014,Keller+2016,Gentry+2017,Gentry+2019}, and more isolated SNeIa \citep{Tang+2009}, as well as CCSNe with progenitors being fast-moving runaway stars \citep[see e.g.,][]{Ceverino&Klypin2009,Kimm&Cen2014,Andersson+2020}. Nonetheless, uncertainties regarding the fraction of runaway stars \citep{Stone1991,Silva&Napiwotzki2011,Eldridge+2011,MaizApellaniz+2018,Renzo+2019,Drew+2021} make their contribution to isolated SNe an unsolved problem. Furthermore, simulations with an explicit treatment of runaway stars find contradicting results. \cite{Andersson+2020} found that runaway stars exploding in low-density gas located in the inter-arm regions of large spiral galaxies result in increased outflow rates. In the dwarf galaxy simulations presented in \citet{Steinwandel+2022}, runaway stars were found to escape the disc of the galaxy, providing thermal energy directly to gas in the circumgalactic medium (CGM). While both these works found runaway stars to play an important role in the galactic scale outflows, \citet{Kim&Ostriker2018} found runaway stars to have negligible effects on these outflows in simulations of stratified $\rm kpc$-sized patches of the ISM \citep[see also][]{Kim+2020a}. Because of significant model variation (e.g. environment, runaway star model, and numerical scheme), no consensus is yet reached for the effect that runaway stars have on feedback physics.

One aim of this work is to study the role that the natal kinematics of individual stars (including walkaway and runaway stars) have on dwarf galaxies, in particular, the role played by the fraction of runaway stars. Dwarf galaxies are both common in the Universe \citep{Sawala+2015,Read+2017,Behroozi+2019}, and they exhibit strong winds relative to their star formation rates \citep{Chisholm+2017,McQuinn+2019}. Furthermore, galactic outflows driven by strong feedback are a necessary component in the $\Lambda$-cold dark matter cosmological model to explain the faint-end of the galaxy-luminosity function \citep{Dekel&Silk1986,Benson+2003}, and the cored density profiles observed in many dwarf galaxies \citep{Moore1994,Teyssier+2013,Read+2016}. Their low escape velocities and relatively large gas contents make them sensitive probes of stellar feedback physics \citep{Rosdahl+2015,Hu+2017,Emerick+2018,Su+2018,Hu2019,Smith+2019,Wheeler+2019,Agertz+2020,Smith+2021}, the stellar initial mass function (IMF) \citep{Smith2021,Prgomet+2021}, and cosmic rays \citep{Dashyan+2020,Farcy+2022,Girichidis+2022}. As numerical laboratories, the small sizes of dwarf galaxies make them less computationally expensive compared to Milky-Way-sized objects, therefore allowing for a large number of simulations at high numerical resolution. In the case of this work, it enables us to run a suite of simulations with a varying fraction of runaway stars, while achieving a resolution high enough to capture important aspects of stellar feedback (e.g. the Sedov-Taylor evolution of SNe, see more details in \sect{numerical_setup}).

Our paper describes our star-by-star model \inferno, as well as presents the theoretical work that motivations our model in Section~\ref{sec:model}, details the numerical set-up and initial conditions (ICs) in Section~\ref{sec:numerical_setup}, and presents the results in Section~\ref{sec:results}. We discuss our results and place our work in a wider context in Section~\ref{sec:discussion}, and finally summarize and conclude in Section~\ref{sec:conclusions}. 

\section{The INFERNO model}\label{sec:model}

\subsection{Star formation, IMF sampling \& initial kinematics}\label{sec:SF_IMF_KICKS}

Following \citet{Andersson+2020,Andersson+2021}, our model incorporates particles representing individual stars to follow stellar motions and feedback for stars above a mass threshold. The threshold is set by a parameter \mstar, and its value determines whether the feedback is calculated for individual stars, or taken as an average over the stellar population below \mstar. Note that \mstar\ can take any value within a given IMF, and while small values employ a more detailed stellar model, it increases the computational cost. Using any predefined IMF, individual stars are stochastically sampled from mass $M_{\rm sf}$ (set as a user-defined parameter; see details in next paragraph). Star formation ensues in each cell with cold ($T<10^4\K$) and dense ($\rho_{\rm g}>500\cc$) gas. At each fine time step, several $M_{\rm sf}$ units of mass can be spawned through a Poisson sampling of a Schmidt-like star formation law,
\begin{equation}\label{eqn:sfr}
    \dot{\rho}_{\rm sf} = \epsilon_{\rm ff}\frac{\rho_{\rm g}}{t_{\rm ff}},
\end{equation}
where $\epsilon_{\rm ff}=0.1$ is the star formation efficiency per free-fall time, and $t_{\rm ff}=\sqrt{3\pi/32{\rm G}\rho_{\rm g}}$ is the local gas free-fall time. 
In the Milky Way, the star formation efficiency per free-fall time is observed at $\sim1\%$ with a large spread \citep[see, e.g.,][]{Krumholz&Tan2007,Lee+2016,Chevance+2022}. Nonetheless, \citet{Grisdale+2017} showed that on $\pc$-scales, a larger value ($\sim10\%$) results in a better match between simulations and observations \citep[see also][]{Grisdale+2018,Grisdale+2019}. We note that the choice of $\epsilon_{\rm ff}$ can affect the properties of the ISM and the outflows, in particular, if set too low \citep[][]{Hu+2022}. Note that in small enough cells, there is not necessarily enough mass $M_{\rm sf}$ to sample individual stars when the density reaches the density threshold for star formation. In such cases, star formation is delayed until enough mass is available.
The population of stars with mass ($m<\mstar$) is traced by one star-particle per star formation event, and can inject feedback based on the model from \citet{Agertz+2013}. In this work, we keep $\mstar$ small enough ($2\Msun$), such that in practice this model is never applied, i.e. stars in this mass range never enter a stellar evolution phase with mass, momentum, or energy ejection.

To sample individual stars from stellar ensembles, we employ the method by \citet{Sormani+2017}, in which the IMF is sampled in predefined mass bins. A detailed description of our implementation can also be found in \citeauthor{Andersson+2020} (\citeyear{Andersson+2020}, see also \citealt{Sormani+2017}). In short, the number of stars in a given stellar mass bin is determined by random number generation from a Poisson distribution with appropriate pre-computed weights. To avoid oversampling, the available mass is sampled consecutively from low to high mass, stopping the process when the available mass is reached\footnote{This model sometimes sufferers from under-sampling the most massive stars, which affects the stellar feedback budget. However, the steepness of the IMF makes this under-sampling rare (handful of times per Gyr).}. To minimize this problem we choose the mass of stellar ensembles to be $M_{\rm sf}=500\Msun$, ensuring a well-sampled IMF \citep{Smith2021}. For this work we use the IMF from \citet{Kroupa2001}, defined as a split power-law function $\xi\propto m^{-\alpha_i}$, with two different mass ranges; $\alpha_1=1.3$ for masses $0.08-0.5\Msun$, and $\alpha_2=2.3$ for masses $0.5-100\Msun$.

At birth, all stellar particles receive the velocity of the gas from which they formed. For individual stars, we give the particles an additional radial velocity with isotropic distribution to model the dynamics which are unresolved in our collisionless simulations. We include two models for this: 1) stars from the same stellar ensemble receive an innate velocity dispersion $\sigma_v$, using random sampling from a Gaussian distribution (referred to as \stir); 2) velocity kicks to simulate walkaway and runaway stars\footnote{Walkaways as stars are typically referred to as stars with peculiar velocities $v<30\kms$, while runaways have $v>30\kms$. We use this convention in our work. These stars originate from either the internal dynamics of star clusters \citep{Poveda+1967} or via binary system breakup due to instantaneous mass loss from companion SNe \citep{Blaauw1961}. Both these scenarios favor more massive stars becoming runaways. The former is due to mass segregation, moving massive stars to the dense center of the cluster, and the latter is due to binary fraction increasing with stellar mass. The kick distribution we apply to escaping stars was estimated from numerical simulations of the first $3\Myr$ of the cluster's evolution \citep[see][for details]{OhKroupa2016}. This does not account for the SNe break-up of binary systems, which is constrained by the time of the first SNe ($\gtrsim3\Myr$). The velocity distribution used for escapers results in $86\%$ walkaways and $14\%$ runaways.}, which overrides the velocity from \stir (referred to as \kick). The \stir model is applied to avoid stars formed at a single instance to remain perfectly overlapping. We emphasize that the \stir model does not entail an accurate treatment of the collisional dynamics on small scales, which are affected by gravitational softening. The \kick model, applied to a fraction $f_{\rm kick}$ of the massive ($>8\Msun$, unless otherwise stated) stars, models walkaway and runaway stars associated with early dynamical interactions in natal star clusters. For this work, we use the inverse power-law distribution $f_v\propto v^{-1.8}$, covering the range $3<v<375\kms$. This is the velocity distribution of stars escaping a $10^{3.5}\Msun$ natal cluster in its first $3\Myr$ of evolution, as modeled by \citet{OhKroupa2016}. This is the same distribution used for massive stars ($>8\Msun$) in \citet{Andersson+2020}, and one of two runaway star models tested by \citet{Steinwandel+2022}.

\subsection{Stellar evolution \& feedback}\label{sec:feedback_model}

\begin{figure}
    \centering
    \includegraphics[width=\linewidth]{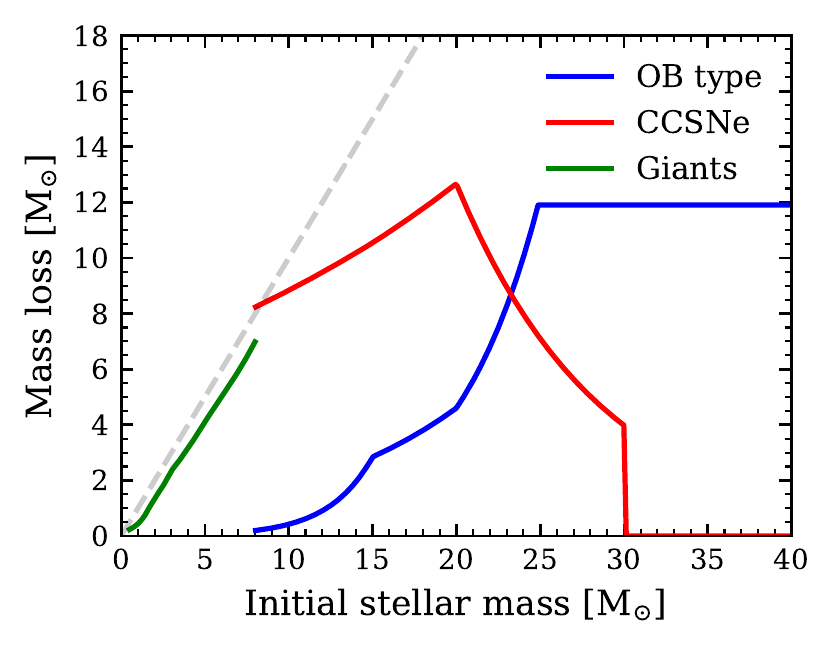}
    \caption{Total mass loss as a function of initial stellar mass shown for the feedback sources considered, denoted in the legend. The grey dashed line shows the equivalence between the two axes. Values are derived by interpolating the results from the NuGrid data sets, and applying the limits constraining the interpolations (see main text for details).}
    \label{fig:mloss}
\end{figure}

\inferno accounts for the injection of energy, momentum, and chemically enriched material, with a model taking the initial mass, metallicity, and age of a given star into consideration. These models apply to different stellar evolutionary stages, and each is described in detail throughout this section. The main factor determining when stars move between evolutionary phases is the main-sequence lifetime. We calculate this using the fitting function from \citet{Raiteri+1996} calibrated to the Padova tracks \citep[][]{Alongi+1993,Bressan+1993,Bertilli+1994}.

The chemical evolution of stars and gas is based on stars inheriting the chemical composition of the gas from which they form, and then injecting chemically enriched material (henceforth referred to as yields). To determine the yield of a given stellar evolution process we use bilinear interpolation of yield tables from \nugrid \citep{Pignatari+2016,Ritter+2018}. This set provides yields for a wide range of stellar masses and metallicities, although we note that there are other yield tables in the literature, with large differences in total yield \citep[see][for a comparison]{Buck+2021}. This method allows us to track up to 80 of the elements in the periodic table, which we describe in more detail in Andersson et al. (in preparation). The stellar evolution models depend only on the total stellar metal mass which we approximate as $M_{\rm Z}=2.09M_{\rm O}+1.06M_{\rm Fe}$, based on Solar mixture \citep{Asplund+2009}.

Similarly to the yields, all mass loss is computed by interpolating the \nugrid tables. \fig{mloss} shows the total mass lost through different feedback channels as a function of the initial mass of a given star. Note that we ensure that the mass expelled by a given star can never result in particles with a negative mass.

\subsubsection{Winds from massive O \& B stars}\label{sec:OBwinds}

The most massive stars ($>8\Msun$) have high enough luminosity to push away material from their surface during the main-sequence phase of their evolution. During this phase, stars launch a fast ($\sim1000\kms$) stellar wind. This wind is driven by the extreme stellar radiation, pushing on the stellar envelope through resonant line absorption \citep{Vink+2015}. Due to its early onset after the formation of a star, this wind can aid the disruption of star-forming clouds, suppresses star formation locally, and affects the clustering of stars \citep[see e.g.,][]{Dale&Bonnell2008,Rosen+2014,Lancaster+2021}.

Our model assumes that all stars in the mass range $8-60\Msun$\footnote{We note that this mass range does not include all B-type stars. For lower mass stars ($<8\Msun$) of this class, we refer to Section \ref{sec:AGB} for details about wind treatment.} launch a wind at a constant velocity of $1000\kms$, for the entire duration of the main sequence. Depending on the stellar mass, the mass loss rates range from roughly $10^{-8}-10^{-6}\Msunyr$. As shown in \fig{mloss}, the mass-loss rate increases non-linearly with stellar mass making extrapolation above the NuGrid upper mass limit ($25\Msun$) sometimes exceeding the initial stellar mass. To avoid this, we assume a constant mass-loss rate for all stars more massive than this limit. This implies that our model likely underestimates the amount of momentum and energy from these winds, although we note that typical IMFs make stars with mass $>25\Msun$ rare.

\begin{figure}
    \centering
    \includegraphics[width=1\linewidth]{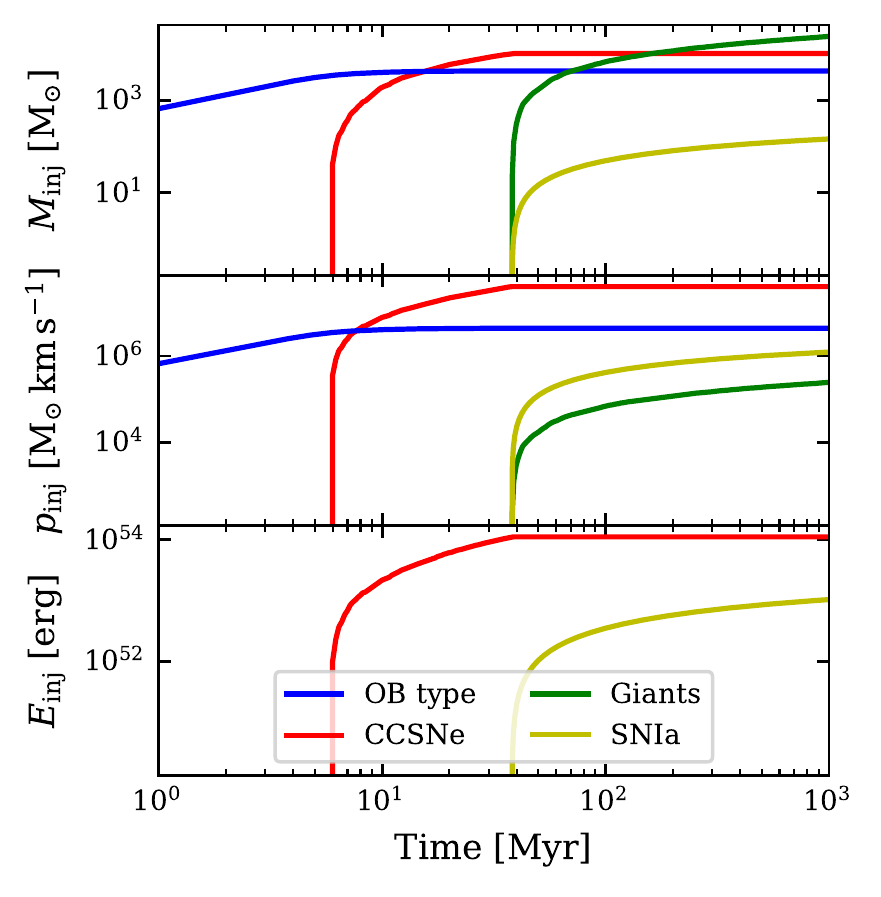}
    \caption{Cumulative sum of the mass (upper), momentum (middle) and thermal energy (lower) injected as a function of time by a $10^5\Msun$ mono-age population of stars. Different sources are distinguished by line color denoted in the figure legend.}
    \label{fig:fbk_inj}
\end{figure}

\subsubsection{Core-collapse supernovae}\label{sec:SNII}

CCSNe results in the instantaneous release of $\sim10^{51}\erg$ of energy, making them a crucial component of any stellar feedback model \citep{McKee&Ostriker1977,Katz1992,Kimm+2015}. The explosion is triggered at the end of the main sequence for massive stars ($\gtrsim8\Msun$), however, the exact mechanism behind the explosion is not fully understood\footnote{The currently favored hypothesis is delayed neutrino-heating, which ejects the outer layers of the stars \citep[see][for a review]{Janka2012}.}. This uncertainty is the often-called {\it islands of explodability} \citep{Janka2012,Zapartas+2021}, stating that many models favor specific ranges in stellar mass to trigger an explosion, with the alternative being the direct collapse to a compact object. Typically, the most massive stars go through the direct collapse channel, however, extremely massive stars ($>100\Msun$) can undergo pair-instability explosions resulting in the complete disruption of the star \citep[see e.g.][]{Fryer+2001}.

Keeping the above complexity in mind, our model assumes that all stars in the mass range $8-30\Msun$ undergo SNe after leaving the main sequence, instantaneously depositing $10^{51}\erg$ of energy, along with chemically enriched material, into its immediate surroundings. The mass expelled during the SNe event is shown by the red line in \fig{mloss}. For stellar masses above this range, we assume that leaving the main sequence results in direct collapse into a black hole, without any injection of energy or enriched material. This implies that the earliest possible injection of energy via SNe occurs $6\Myr$ after star formation (see \fig{fbk_inj}).

\subsubsection{Stellar winds from giant stars}\label{sec:AGB}

Stars more massive than $0.5\Msun$ enter a giant phase for a short period after leaving the main sequence unless the star undergoes SNe before this. In this phase, energy is mostly generated through hot-bottom burning in convective shells exterior to the stellar core, periodically supplying the core with fuel giving rise to explosive burning \citep[see][for a review]{Hofner&Olofsson2018}. These surges in energy (often called thermal pulses) drive a stellar wind with mass loss rates in the range $10^{-8}-10^{-4}\Msunyr$ at velocities $\approx10\kms$ \citep[see e.g.][]{Schoier&Olofsson2001,Olofsson+2002,GonzalezDelgado+2003,Ramstedt+2009,Eriksson2014}. Although this wind makes up only a small fraction of the stellar feedback energy budget, it is crucial for the chemical enrichment of the ISM. A source of uncertainty in stellar evolution models with regards to giant stars is the intermediate phase ($7.5-9\Msun$) between evolving into a white dwarf or CCSNe \citep{Poelarends+2008,Doherty+2017}. After leaving the main sequence, these stars are massive enough to ignite carbon burning in their core, resulting in a large number of thermal pulses giving rise to a super asymptotic giant branch phase. During this phase, material fueled to the core can result in its mass exceeding the Chandrasekhar mass, leading to the core explosion. 

Our model assumes that all stars in the mass range $0.5-8\Msun$ enter a post-main-sequence phase, during which a stellar wind is expelled. The wind is injected as a source of momentum at a constant mass loss rate of $10^{-5}\Msunyr$ with a velocity of $10\kms$. The duration of this phase is set by the total mass lost (green line in \fig{mloss}), computed from the \nugrid tables, i.e. winds are expelled until no more mass is available, in which case the star is considered to have become a white dwarf. The resulting initial-final mass relation roughly matches that in \citet{Cummings+2016}.

\subsubsection{Type Ia supernovae}\label{sec:Ia}

SNeIa are essential for the chemical evolution of galaxies as they are a source of Fe-peak elements, with some contribution to $\alpha$ \citep[see e.g.,][]{Seitenzahl+2013,Kobayashi+2020}. Although their origin is still not fully understood, mass transfer to a degeneracy-supported object in a binary system seems ubiquitous to models, with a near-Chandrasekhar-mass white dwarf primary being the most favorable candidate \citep{Bloom+2012}. Due to their uncertain origins, empirical models assuming delay-time distributions weighted by cosmic star formation histories are often used for modeling SNeIa rates \citep[see e.g.,][]{Mannucci+2006,Maoz+2014,Maoz&Graur2017}. 

Our model incorporates the field normalized delay-time distributions from \citet{Maoz&Graur2017}, giving a SNeIa rate per unit mass
\begin{equation}\label{eqn:Ia_model}
    n_{\rm Ia}=I_{\rm Ia}\left(\frac{t}{\Gyr}\right)^{-1.12}\Delta t,\quad t>t_{\rm Ia},
\end{equation}
assuming a delay time $t_{\rm Ia}=38\Myr$ (main sequence lifetime of $8\Msun$ star), and normalization $I_{\rm Ia}=2.6\times10^{-13}\yr^{-1}\Msun^{-1}$. Because of the uncertainty regarding progenitor\footnote{Note that these rates do not assume a progenitor, however, our chemical yield model does. SNeIa yields from \citet{Seitenzahl+2013} assumes a Chandrasekhar-mass delayed-detonation scenario.}, as well as a missing tracer for binary stars in our model, we use the particles tracing the unresolved stellar component to determine possible locations of SNeIa. To compute the number of SNeIa, each star particle representing unresolved stars stores the total mass of coeval stars and uses it to normalize $n_{\rm Ia}$ for a given star particle age. This number ($\ll1$) is used to determine the probability of an event, ultimately sampling discrete SNIa. Each explosion releases $10^{51}\erg$ of energy and $1.4\Msun$ of mass into its immediate surrounding.

\fig{fbk_inj} summarizes the feedback budget of our model, showing the cumulative mass, momentum, and energy which is injected into the surroundings of a $10^5\Msun$ mono-age population of stars over $1\Gyr$. With the exception of winds from OB-type stars, the onset of feedback from the different sources is determined by the main-sequence lifetime of the most massive star in the relevant population of stars. The range of timescales for this onset highlights the importance of including a multitude of feedback sources, as this affects both how star formation proceeds locally, and shapes the environment for subsequent feedback. 

\section{Numerical setup and model implementation}\label{sec:numerical_setup}

\begin{figure}
    \centering
    \includegraphics[width=0.95\linewidth]{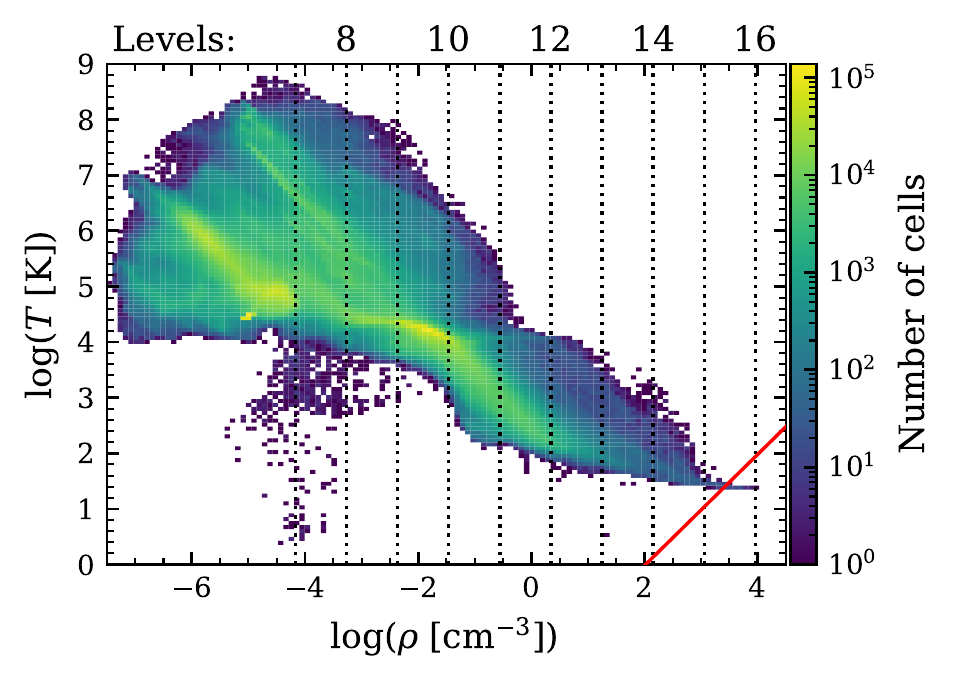}
    \caption{An example showing the number of cells at different densities and temperatures in one of our simulations. Vertical dotted lines show the density where refinement is triggered from the indicated refinement level to the next. The red line shows the density and temperature where the thermal Jeans length is equal to the resolution at the highest refinement level ($1.5\pc$), however, see discussion in the main text. The data shown spans the entire simulation box, and thus includes spurious effects from the box boundaries (e.g., the few cells at $\log(\rho)\sim-4$ at $\log(T)\lesssim4$). Note that the levels indicated only reflect refinement based on cell mass ($100\Msun$), and additional refinement criteria are applied as well (e.g., an average of 8 particles per cell).}
    \label{fig:resolution}
\end{figure}

\inferno is implemented in the adaptive-mesh-refinement (AMR) and $N$-body code \ramses \citep[][]{Teyssier2002}. \ramses evolves the gas by solving the fluid equations on a refinement grid with a second-order unsplit Godunov method, assuming an ideal mono-atomic gas with an adiabatic index of $5/3$. 
The cooling module applied combines equilibrium thermochemistry of hydrogen and helium \citep[][]{Courty&Alimi2004,Rosdahl+2013}, metal line cooling rates computed with \cloudy \citep[][]{Ferland+1998}, and a uniform UV background \citep[][]{Haardt&Madau1996} including an on-the-fly self-shielding model \citep[][]{Aubert&Teyssier2010,Rosdahl&Blaizot2012}. The equilibrium chemistry of hydrogen and helium considers photoionization, collisional ionization and excitation, recombination, free-free emission, Compton cooling and heating, and dielectronic recombination. For a more detailed discussion, see \citet{Rey+2020}. We limit gas cooling down to a minimum temperature of $1\K$. 
The dynamics of stars and dark matter are tracked using collisionless particles, whose contribution to the gravitational potential is added to the AMR grid with the cloud-in-cell particle-mesh method. The forces are calculated by solving the Poisson equation with a multi-grid method \citep{Guillet&Teyssier2011}. A quasi-Lagrangian refinement strategy ensures roughly 8 particles in each cell, which reduces discreteness effects \citep{Romeo+2008}. Furthermore, cells are split into 8 new cells, using a refinement mass criterion of $8\times100\Msun$. We limit cell-splitting to $16$ levels of refinement, providing a spatial resolution limit of $\sim1.5\pc$ for our simulations, which are set up in a box with $100\kpc$ side length. \fig{resolution} highlight the frequency of cells at the different resolution (refinement levels) in a phase diagram ($\rho$ versus $T$) taken as a representative simulation output. We do not consider the Jeans criterion in our refinement strategy. \citet{Hopkins+2018} showed that the ISM is primarily supported by turbulence rather than thermal pressure, even down to cloud scales. This fact alleviates the resolution requirement set by, e.g., \citet{Truelove+1997}  (note that the red line in \fig{resolution} shows the density and temperature at which the thermal Jeans length is equal to our smallest cell size). Thus we argue that the effective Jeans length (thermal plus turbulent) is likely resolved in our simulations, with cold gas at high densities being treated by the star formation recipe (see also discussion in \citealt{vanDonkelaar2022}).

\begin{figure*}
    \centering
    \includegraphics[width=0.9\linewidth]{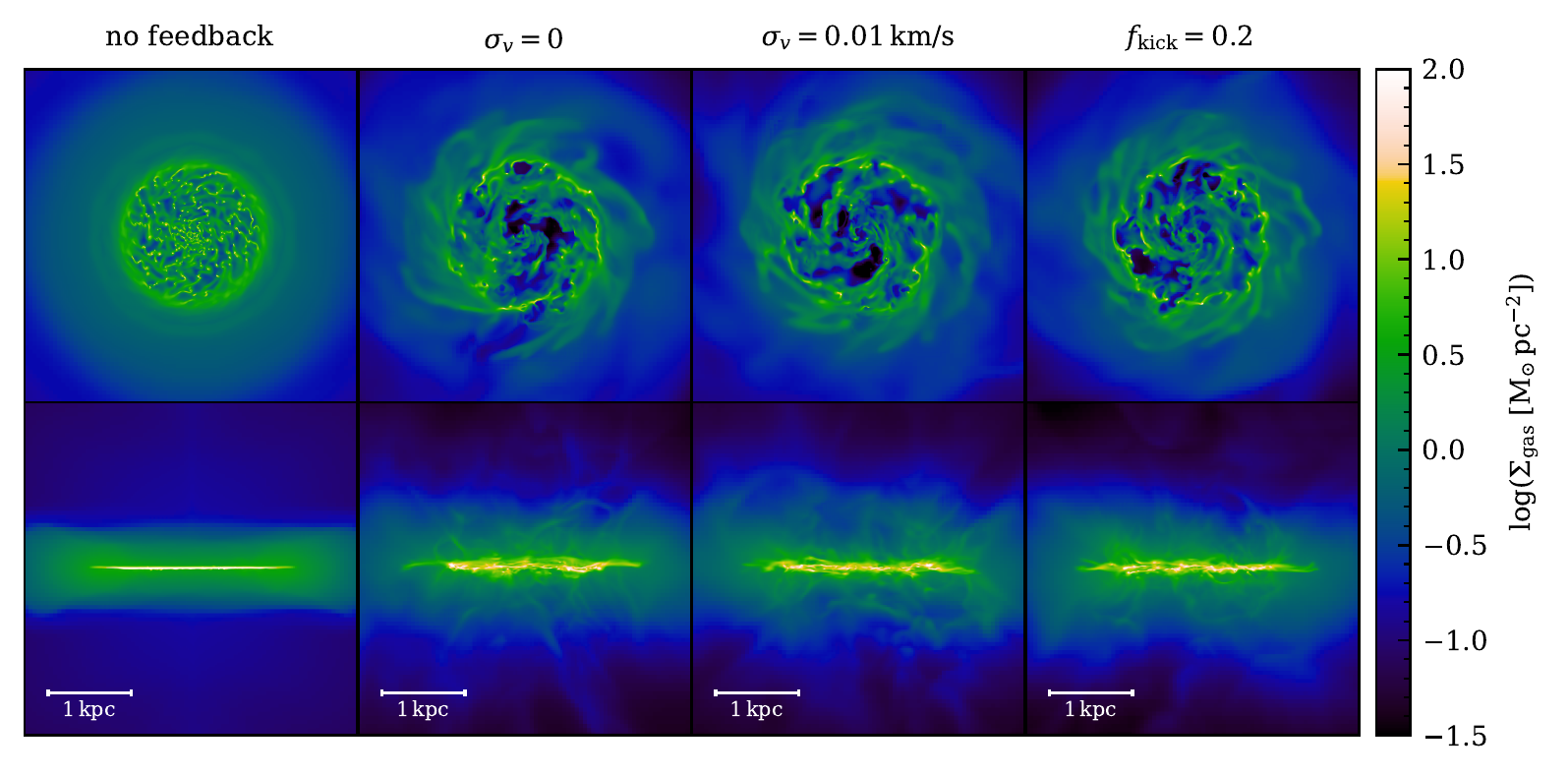}
    \caption{Projected gas density of the simulations studied in this work, shown in face-on projection on the top row and edge-on projection on the bottom. The snapshots shown are at $t=400\Myr$, and all snapshots for a given simulation in the time-span $300-500\Myr$ are similar, with the exception of transient events, such as super-bubble outbreaks.}
    \label{fig:rhomap}
\end{figure*}

We employ \inferno on a dwarf galaxy to study how efficiently stellar feedback drives outflows. The simulated galaxy is an analog of the Wolf–Lundmark–Melotte (WLM) galaxy with a gas mass $M_{\rm g,disc}\approx7\times10^{7}\Msun$, an initial stellar disc with mass $M_{\rm s, disc}=10^{7}\Msun$ and a dark matter halo with mass $M_{\rm vir}=10^{10}\Msun$. The latter two are comprised of $12.5\Msun$ stellar particles and $1650\Msun$ dark matter particles. We consider the initial stellar component only as a mass component (i.e. with no contribution to feedback or enrichment). The initial disc, comprised of gas and stars, has an exponential radial density profile with a scale length of $1.1\kpc$. The vertical gas distribution is set in accordance with hydrostatic equilibrium at an initial temperature of $10^{4}\K$, while the vertical distribution of stars is initialized with a Gaussian distribution with a scale height of $0.7\kpc$. Initially, the gas disc has a metallicity of $0.1\Zsun$. The dark matter profile matches an NFW profile \citep{NFW1996} with a spin parameter $\lambda=0.04$ and concentration parameter $c=15$. The ICs were generated using {\sc MakeDiscGalaxy} \citep{Springel2005}) and mapped onto the AMR grid using the cloud-in-cell method. These generated ICs do not fill the full extent of our simulated box, hence cells without assigned properties are initialized with a density of $10^{-5}\cc$, a metallicity of $0.001\Zsun$, and a temperature of $3\times10^4\K$. These ICs are almost identical to those in \cite{Smith+2021}.

Since we do not consider feedback processes from the stars included in the ICs, the initial gas support is purely thermal. This energy support is quickly radiated away resulting in a sudden collapse and star formation burst, which is typical for galaxies simulated in isolated boxes. To mitigate this effect we start the simulation without gas cooling and then ramp it up exponentially (formally we scale the internal energy sink responsible for cooling by $(t/t_0)^5$, effectively re-scaling the cooling rate) over the first $t_0=100\Myr$. This method allows for a calm initialization of the galaxy. We do not include this transient in any of our result figures.

The stellar feedback model injects energy, momentum, and chemically enriched material at each fine time step (i.e. between the time integration of each refinement level). Every timestep we loop through all stars and inject the relevant feedback quantities into the oct closest to the star particle (8 neighboring cells), updating the density, velocity, and pressure of each cell. Momentum is added isotropically. If a star enters a new evolutionary stage during a timestep (which affects the feedback model), we adapt the calculation to only cover the part of the timestep during which stellar feedback is active. Furthermore, two safety criteria (a maximum advection velocity of $6000\kms$, and a maximum temperature of $10^{9}\K$) are employed to ensure the stability of the hydrodynamics solver.

Because the resolution is limited (specifically in low-density gas by the AMR prescription) the momentum buildup in the quasi-energy-conserving stage of SNe explosions is not always captured. To handle this problem, we first calculate the radius $r_{\rm ST}$ of the blast-wave when it transitions from energy conserving to momentum conserving \citep[i.e., from the Sedov-Taylor phase to the often called snowplow phase,][]{Sedov1959,Taylor1950}. If this radius is not resolved by at least 6 cells we inject the terminal momentum $p_{\rm ST}$ that would have built up during the energy conserving stage. We compute the cooling radius from $r_{\rm ST}=30\,E_{51}^{7/17}\,\rho_{\rm g}^{-7/17}\,Z_{\rm g}^{-0.2}\pc$, where $E_{51}$ is energy in units of $10^{51}\erg$, $\rho_{\rm g}$ is cell density in units $\cc$, and $Z_{\rm g}$ is metallicity in solar value. This follows from the analytical blast-wave solution \citep[][]{Blondin+1998}, to which we have added a metallicity scaling calibrated to our cooling function \citep[][]{Thornton+1998}. Similarity, the terminal momentum is calculated from $p_{\rm ST}=2.95\times10^{5}\,E_{51}^{16/17}\,\rho_{\rm g}^{-2/17}\,Z_{\rm g}^{-0.2}\Msun\kms$, where we have adjusted the scaling following \citet{Kim&Ostriker2015}. Based on the blast wave criterion, roughly $5\%$ of SNe are unresolved in our simulations. In recent work, \citet{Hu2019} suggests that the injection of momentum does not accurately capture the evolution of a SNe. Although unexplored for the type of code employed here, we caution the reader that $5\%$ of our SNe might underestimate the amount of energy incorporated into the ISM. We leave a detailed exploration of this for future work, but see Appendix \ref{sec:convergence_test}.


\section{Results}\label{sec:results}

\begin{figure}
    \centering
    \includegraphics[width=1\linewidth]{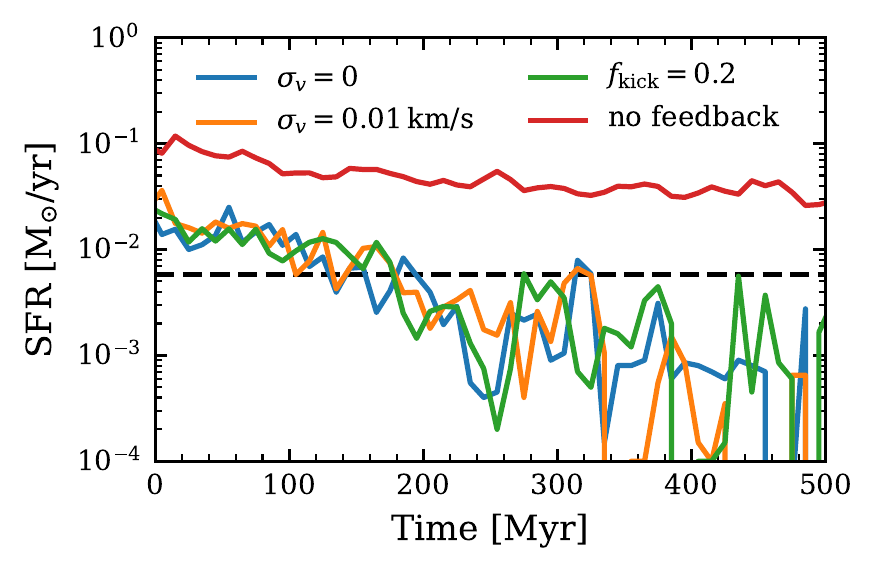}
    \caption{Star formation rate as a function of time for our simulations, computed by summing the stellar mass in $10\Myr$ age bins. The dashed horizontal line shows the observed star formation rate of WLM \citep{Karachentsev+2013}. With the inclusion of stellar feedback, star formation is significantly reduced as a result of the loss of cold gas.}
    \label{fig:sfr}
\end{figure}

\begin{figure*}
    \centering
    \includegraphics[width=1\linewidth]{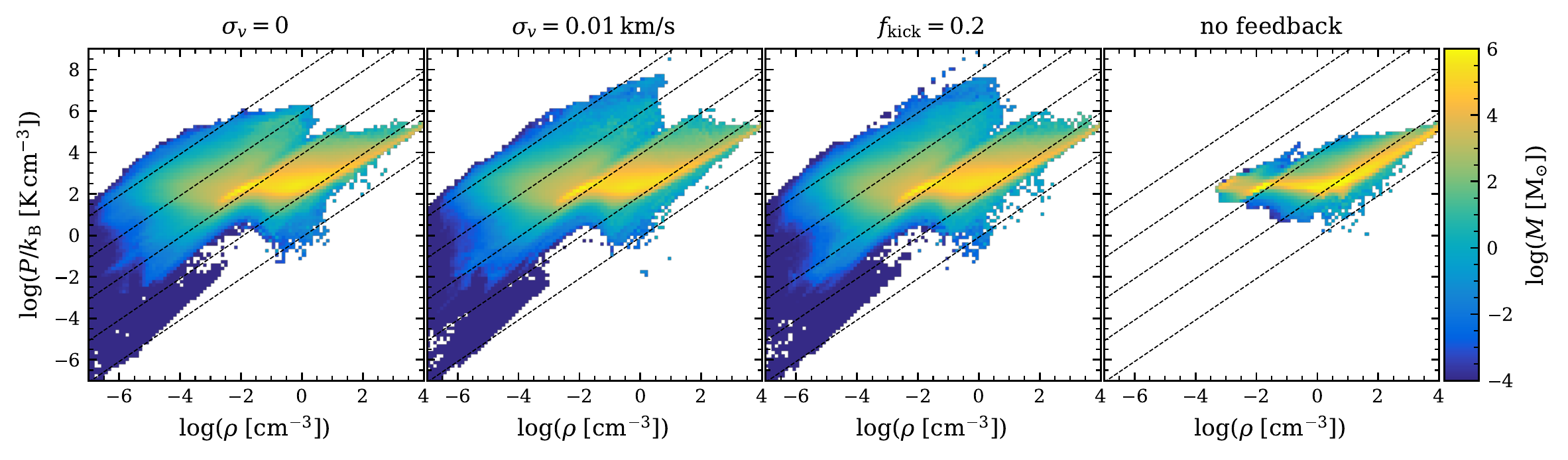}
    \caption{Pressure at different gas densities in the disc of our simulations (denoted by the title of each panel). These quantities are measured in each cell for all outputs and plotted as time-averaged ($300-500\Myr$), mass-weighted 2D-histogram. The dotted lines show temperatures $10^0$, $10^2$, $10^4$, $10^6$, and $10^8\K$ from the bottom right to the top left, computed from the ideal gas law.}
    \label{fig:rhoP}
\end{figure*}

\begin{figure}
    \centering
    \includegraphics[width=1\linewidth]{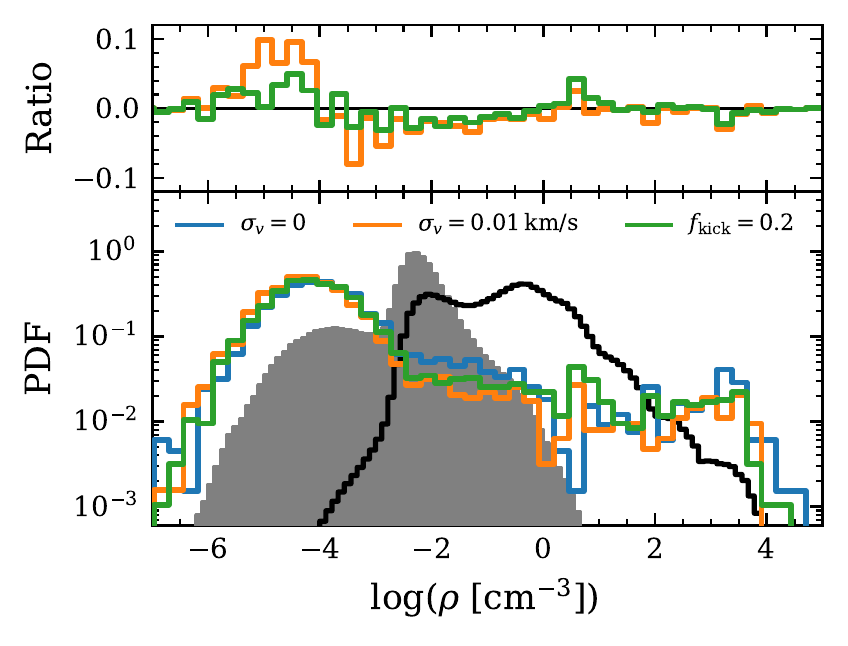}
    \caption{Probability distribution of gas density, showing volume (mass) weighted distributions for $f_{\rm kick}=0.2$ in the filled grey histogram (thick black line), and colored lines (different models labeled in legend) showing the distribution of densities where CCSNe exploded. Note that all models with feedback have similar volume and mass-weighted distributions. All results are taken as time averages for the final $200\Myr$ of each run. The top inset shows the ratio of the distributions between the two models including natal stellar velocities (distinguished by the same color as in the bottom plot) and the $\sigma_v=0$ model.}
    \label{fig:rhoSNe}
\end{figure}

As detailed in \sect{intro}, the aims of this work are (i) verifying that \inferno produces realistic ISM conditions for galaxy evolution, (ii) exploring the physics of outflows in a dwarf galaxy, and (iii) investigating how these outflows are affected by natal stellar kinematics. For the latter, we compare the results of 8 dwarf galaxy simulations with identical ICs, but with different natal velocity distributions. In addition, a ninth simulation (\nofb) serves as an example of not including stellar feedback. To maintain clarity, the main body of this work includes the detailed analysis of 4 simulations: 1) \nofb, with no energy or momentum injection from stellar feedback sources; 2) $\sigma_v=0$, with neither \stir or \kick applied; 3) $\sigma_v=0.01\kms$, with \stir applied; 4) $f_{\rm kick}=0.2$, with $\sigma_v=0.01\kms$ \stir and \kick applied. The full suite of simulations are shown in Appendix \ref{sec:all_runs}, where we divided them into \stir models ($f_{\rm kick}=0$), and \kick models ($f_{\rm kick}>0$). The \nofb simulation has a $\sigma_v=0.01\kms$ \stir applied. The choice of $f_{\rm kick}=0.2$ is motivated by the cluster escape fractions ranging from $10$ to $30\%$ for massive stars, as found in \citet{OhKroupa2016}. After the initial relaxation ($200\Myr$) we follow the evolution for $500\Myr$, covering a few orbital times. Our analysis only concerns the final $500\Myr$ of evolution.

\subsection{Effect on interstellar medium}\label{sec:disc_ISM}

With the exception of \nofb, the visual appearances of the gas properties in our simulations are similar (see e.g., gas density in \fig{rhomap}). For \nofb, the absence of energy and momentum sources results in a cold and fragmented disc. In contrast, the inclusion of stellar feedback significantly reduces the number of clouds and creates hot low-density voids in between the gaseous spiral structure. Furthermore, feedback drives gas out of the galaxy, generating a complex gas structure above and below the disc. This inner circumgalactic medium (CGM) is similar in all feedback models, regardless of the natal kick model. This is unlike those found in \citet{Steinwandel+2022}, as well as results for more massive systems, which have been shown to be strongly affected by the inclusion of runaway stars (\citealt{Ceverino&Klypin2009,Andersson+2020}, but see \citealt{Kim&Ostriker2018,Kim+2020a}). We discuss this further in \sect{discussion}.

Feedback leads to a lower star formation rate (SFR), quantified in \fig{sfr}. At early times ($t<200\Myr$), the lack of significant inflows causes the gas content to reduce over time, with an accompanying decrease in SFR in all simulations. For simulations including feedback, the SFR flattens after this period, as fountain flows are starting to regulate the supply of gas to the galaxy. In the first $300\Myr$ of these simulations, the total gas mass fraction $f_g=M_{\rm g}/(M_{\rm g}+M_{\star})$ is reduced by $\sim10\%$, while that of only cold ($T<10^4\K$) gas is reduced by $\sim20\%$. From this, it is clear that feedback reduces the amount of gas available for star formation, in part due to outflows which we focus on in \sect{outflows}. In the simulations including feedback, the SFR is compatible with that observed in the WLM galaxy ($\sim6\times10^{-3}\Msunyr$, \citealt{Karachentsev+2013}) at $\sim200\Myr$. The galaxies evolve with periodic variations in the SFR, on average lying below the observed rate by a factor of few after $200\Myr$. The periodicity arise from bursts in stellar feedback act in response to the periods of high star formation. For example, in the case of $\sigma_v=0.01\kms$, a complete shutdown of SFR occurs between $430$ and $470\Myr$. 

\begin{figure*}
    \centering
    \includegraphics[width=0.75\linewidth]{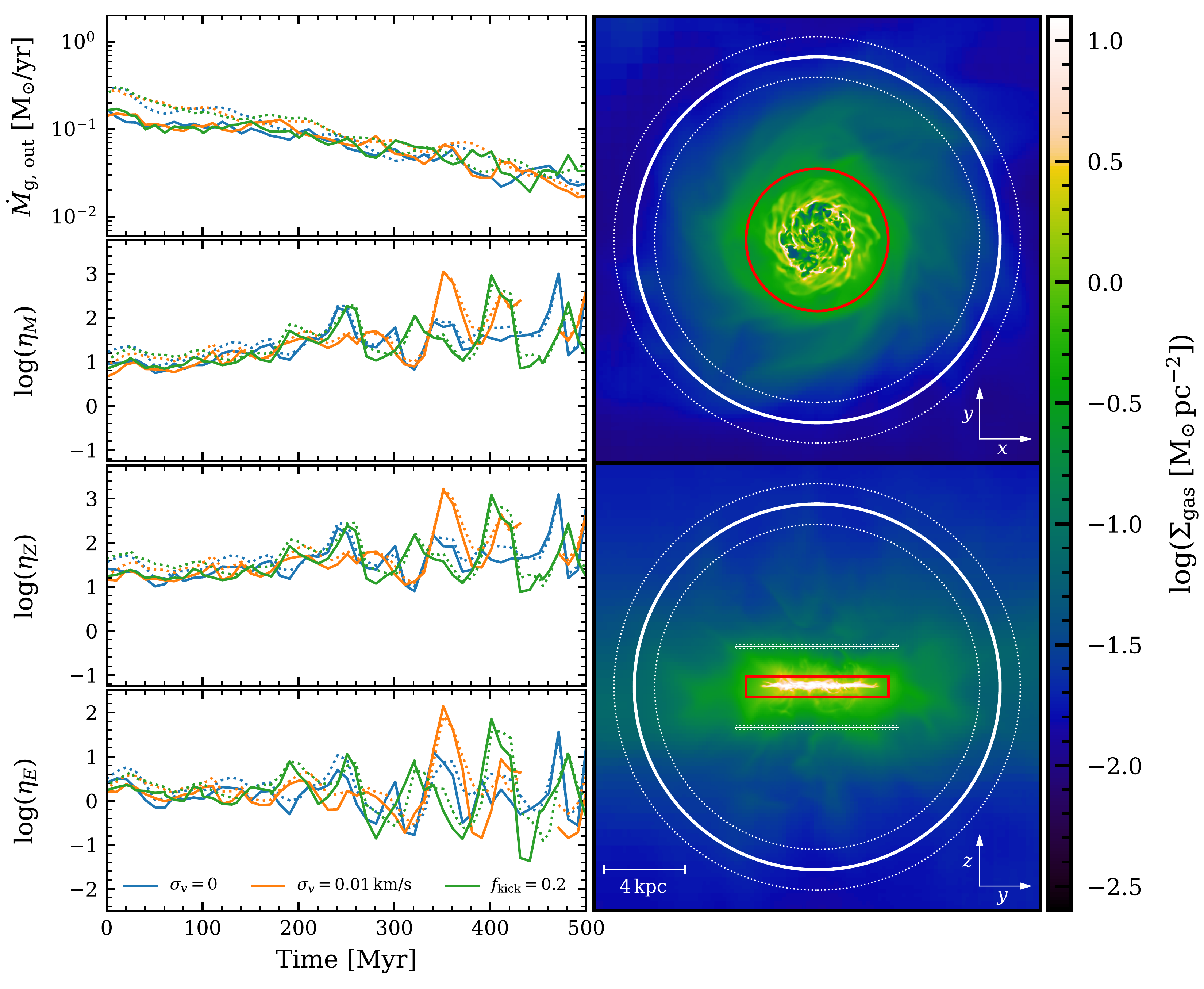}
    \caption{{\it Left:} Mass outflow rate (top) and loading factor of mass (top center), metal (bottom center), and energy (bottom) as a function of time for the simulations including feedback with different natal kinematic models as labeled by the legend in the bottom plot. The filled (dotted) lines show measurements through the \launching (\innercgm) interface. {\it Right:} Projected gas density of the $f_{\rm kick}=0.2$ simulation in a $22\kpc$ view, displaying the placement of the \launching and \innercgm interfaces (white filled and dotted lines). Material encapsulated by the red lines is considered disc material.}
    \label{fig:outflows}
\end{figure*}

In addition to suppressing the SFR, stellar feedback generates an over-pressured hot phase in the ISM and large low-density bubbles. \fig{rhoP} shows this highly multi-phase gas structure of the disc (defined as a cylinder with a radius of $3.5\kpc$ and height of $1\kpc$, as outlined with red lines in the right plots of \fig{outflows}). The pressure $P$ of simulations including feedback spans several orders of magnitude ($P/k_{\rm B}\sim1-10^{6}\K\cc$ at $\rho=1\cc$, where $k_{\rm B}$ is the Boltzmann constant). The majority of high-pressure gas is generated by clustered feedback and the natal stellar velocity model plays only a minor role. The clustered nature stems from vigorous star formation in dense clouds and proceeds until halted by the onset of the first CCSNe ($6\Myr$). Before this disruption of the star-forming clouds, the gas collapse is suppressed by stellar winds.

\fig{rhoSNe} compares the average probability density function (PDF) of the gas densities for the simulations. For a given coeval stellar population, the first CCSNe typically explodes in dense gas ($\rho\sim10^{4}\cc$), rapidly building up a low-density bubble ($10^{-6}<\rho<10^{-3}\cc$) for subsequent CCSNe. During the buildup of the bubble, CCSNe explodes in intermediate gas densities ($10^{-2}\lesssim\rho\lesssim10^{2}\cc$). In the top panel, we show the ratio between the explosion densities of the two simulations including natal kinematic models and the $\sigma_v=0$ model. Surprisingly, $\sigma_v=0.01\kms$ deviates the most from $\sigma_v=0$ (clearly visible at $\rho=10^{-5}\cc$), although the difference is small and subject to stochasticity between measurements. The minor role of runaway stars in determining the explosion density distribution is due to their rarity in comparison to non-runaway stars. For our galaxy, the effect that stellar feedback has on the gas dynamics is completely dominated by the clustered CCSNe. This is apparent in the outflows, which we explore in the following section.

\subsection{Outflows and inner CGM}\label{sec:outflows}

The energy supplied by our feedback model translates into a galactic wind, resulting in large amounts of gas being pushed out of the galaxy. Significant amounts of gas return in galaxy scale fountain flows, while the rest is accelerated to outflow velocities $v_{\rm out}$ exceeding the escape velocity $v_{\rm esc}$ (described in more detail in \sect{tave_wind}). Here we explore the interplay between the outflowing gas and the inner parts of the CGM, focusing on the final $200\Myr$ of the simulations. We no longer consider any results from the \nofb model, due to its inability to generate outflows.

We measure the properties of outflowing gas at two interfaces located outside the galaxy. Their location (white filled) and extent (white dotted) are displayed on top of projected density maps of the $f_{\rm kick}=0.2$ simulation in \fig{outflows}. We refer to the interface located close to the disc as \launching, and the spherical shell interface, which encapsulates the inner CGM, as \innercgm. Note that these definitions vary in the literature, and that outflow properties can depend on how these are defined. We measure the properties of the outflowing gas by summing a quantity $q$ multiplied by the gas velocity $v$, considering only cells with outward moving gas within a given region.\footnote{In this work we mainly consider mass $m_i$, metal mass $Z_im_i$, and total energy $m_i(v_i^2/2+c_{s,i}^2/(\gamma-1))$ outflows, substituting $q_i$ with these terms when applicable. Throughout the paper, we take $Z_i$ to be the metal mass fraction, $c_s$ to be thermal sound speed, and $\gamma=5/3$ to be the adiabatic index. Note that kinetic energy refers to the first term in the total energy sum, while the second term is thermal energy.} Variables indexed by $i$ refer to their value in individual cells. For the \launching interface, this is formally calculated following 
\begin{equation}\label{eqn:qdot_z}
    \frac{{\rm d}q_z}{{\rm d}t} = \frac{1}{\Delta z}\sum_{i} q_{i}\,|v_{z,i}|,\ {\rm for}\
\begin{cases}
v_{z,i} > 0,\ {\rm if}\ z_i>0\\
v_{z,i} < 0,\ {\rm if}\ z_i<0
\end{cases},
\end{equation}
where the sum runs over cells ($i$) in a cylindrical slab with thickness $\Delta z=0.1\kpc$ placed $\pm1\kpc$ from the disc mid-plane. For the \innercgm interface, we use 
\begin{equation}\label{eqn:qdot_r}
    \frac{{\rm d}q_r}{{\rm d}t} = \frac{1}{\Delta r}\sum_{i} q_{i}\,v_{r,i},\ {\rm for}\ v_{r,i}=\mathbf{v}_{i}\cdot\hat{\mathbf{r}}>0
\end{equation}
where the sum runs over cells in a spherical shell with a radius of $9\kpc$ (roughly equal to $20\%$ of the virial radius) and a thickness $\Delta r=2\kpc$. The coordinate system has its origin at the center of the galaxy and its directions are indicated in the density maps to the right in  \fig{outflows}. In Equation \ref{eqn:qdot_r}, $\mathbf{v}_i$ is the velocity vector, and $\hat{\mathbf{r}}$ is the radial unit vector.

\begin{figure*}
    \centering
    \includegraphics[width=0.7\linewidth]{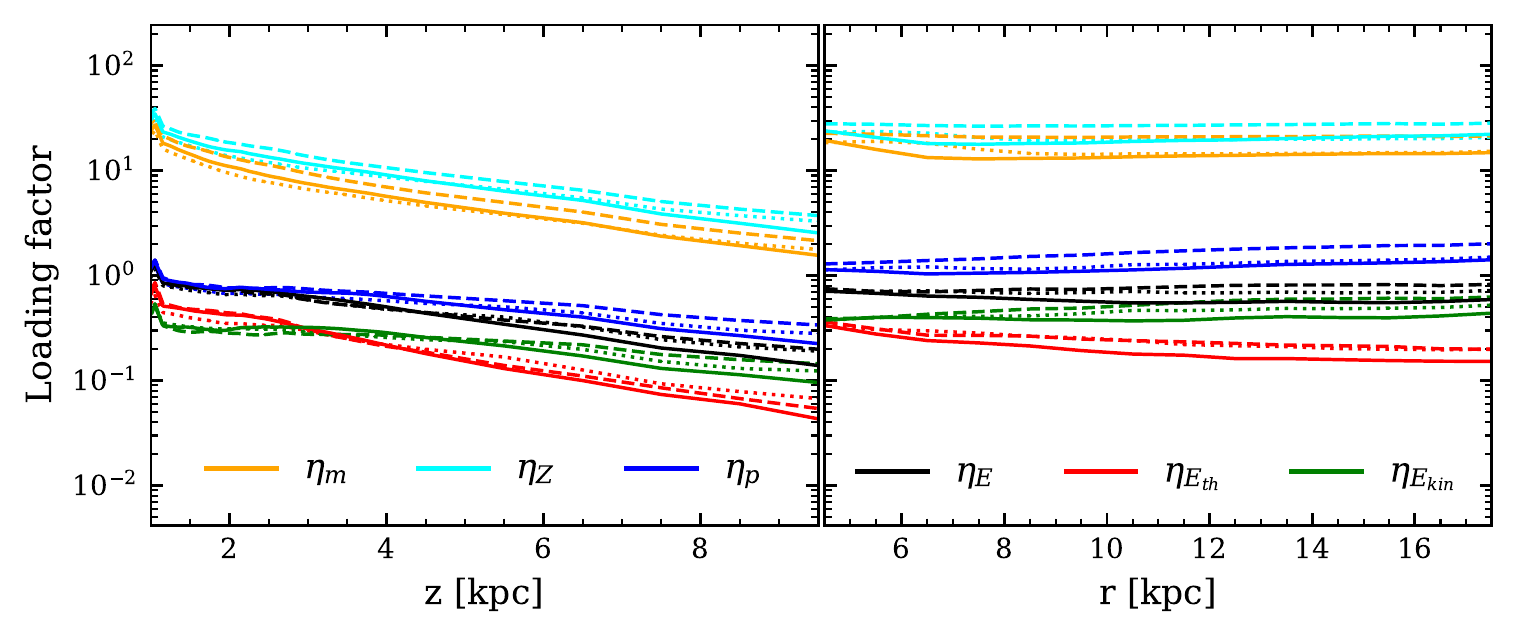}
    \caption{Loading factors of mass (orange), metal (cyan),  momentum (blue), total energy (black), thermal energy (red), and kinetic energy (green) as a function of vertical distance to the left and radial distance to the right. We measure vertical outflow using slabs of the same size as that in \fig{outflows} but increasing the thickness to $1\kpc$ above $4\kpc$. Radial outflows are measured as in \fig{outflows}, but for different radii. The lines show the mean value of all outputs in the last $200\Myr$ (see text for details). $f_{\rm kick}=0.2$ is shown by the filled lines, while the thin dotted (dashed) lines show the results for $\sigma_v=0$ ($\sigma_v=0.01\kms$).}
    \label{fig:eta_prof}
\end{figure*}

The top left plot in \fig{outflows} shows the mass outflow rate as a function of time\footnote{There is some ambiguity in how one defines the vertical outflows, and in our case, we chose to compute the outflows in the two slabs independently and then sum them. For modeling purposes, the total outflow through both slabs is most useful, as it captures the mass loss from the disc. Observationally, the outflow measurements are typically limited to a single cone and then multiplied by a factor of two (assuming symmetry), in order to capture the total outflow rate \citep[see, e.g.,][]{Schroetter2019}. We find little to no difference between the two slabs, justifying this assumption.}. The remaining three left plots show the time evolution of the logarithm of mass, metal, and energy loading factors, which we define as 
\begin{equation}\label{eqn:loading_factor}
    \eta_{M} = \frac{\dot{M}_{\rm g}}{\rm SFR},\quad \eta_{Z} = \frac{\dot{M}_{\rm Z}}{Z_{\rm g}\cdot{\rm SFR}},\quad \eta_{E} = \frac{\dot{E}}{\xi_{\rm SN} \cdot{\rm SFR}}
\end{equation}
respectively, where $Z_g$ is the gas metallicity of the disc, and $\xi_{\rm SN}=4.89\times10^5\,{\rm km}^2\,{\rm s}^2$ is the average energy injected by CCSNe from stellar populations with a fully sampled \cite{Kroupa2001} IMF assuming $10^{51}\erg$ per CCSNe \citep[c.f.,][]{Kim&Ostriker2017,Smith+2021}.

The gas mass ejected in outflows exceeds the mass consumed in star formation by up to two orders of magnitude. This is the case for all three models, which all show outflow rates of similar average values. Furthermore, the values are similar both at the \launching and the \innercgm interfaces. As noted in the previous paragraph, wind properties are in general sensitive to where they are measured and here their similar values are coincidental, as the vertical placement of the \launching interface affects the value measured. Surprisingly, this is not the case for the \innercgm which we discuss in more detail in \sect{tave_wind}.

The mass loading factor increases in the first $200\Myr$, and then reach values that fluctuate between $\sim10-1000$, independently of the natal stellar velocity model. This is also the case for the metal loading factor, although its value slightly exceeds $\eta_M$. The energy loading factors do not show an initial increase but display similar fluctuations around the same time. As with the other loading factors, these fluctuations grow significantly stronger at later times, resulting in values of $\eta_E$ in the range $\sim0.1-100$. These fluctuations are the result of variations in SFR, and the outflow properties remain more stable (see e.g. mass outflow rate in the top left plot of \fig{outflows}).

In a broader context, high mass loading factor ($10-100$) for low mass galaxies are typically required by semi-analytical models \citep[see e.g.,][]{Benson+2003,SomervilleDave2015} and large volume simulations to match observed galaxy scaling relations \citep[e.g,][]{Mitchell+2020}. Metal loading factors are typically found to be of the same order or in excess of the mass loading factor \citep{Yates+2021}, as we also find here. \inferno captures these heavily mass and metal-loaded winds without fine-tuning any feedback parameters. Comparisons of energy loading factors are more difficult since these depend strongly on the details of the feedback model, as well as the cooling and structure of the CGM. In our case, the absence of a cosmological environment, and $\eta_E$ around unity gives rise to a CGM with total energy set by the stellar feedback budget. Compared to studies of outflows with similar feedback model and galaxy \citep[e.g.,][]{Smith+2021,Steinwandel+2022} our values of $\eta_E$ are high, which we discuss further in \sect{discuss_comparison}. Our loading factors roughly match those in observed galaxies \citep{Chisholm+2017,Chisholm+2018}, although it should be noted that completeness issues and differences in the geometrical definition of where outflows are measured make loading factors notoriously difficult to estimate, in particular for dwarf galaxies \citep[see][for a review]{Collins&Read2022}. For mass and energy loading, \citet{Chisholm+2017} accounts only for the photo-ionized gas, which does not necessarily capture the entire outflow (a notion returned to later). Furthermore, because of the strong temporal fluctuations we find in our simulation, a better comparison would be to investigate if the range of loading factors in our simulations matches the scatter in observations. However, such a comparison would necessitate more observational data points for the galaxy mass range we consider.

\begin{figure}
    \centering
    \includegraphics[width=0.85\linewidth]{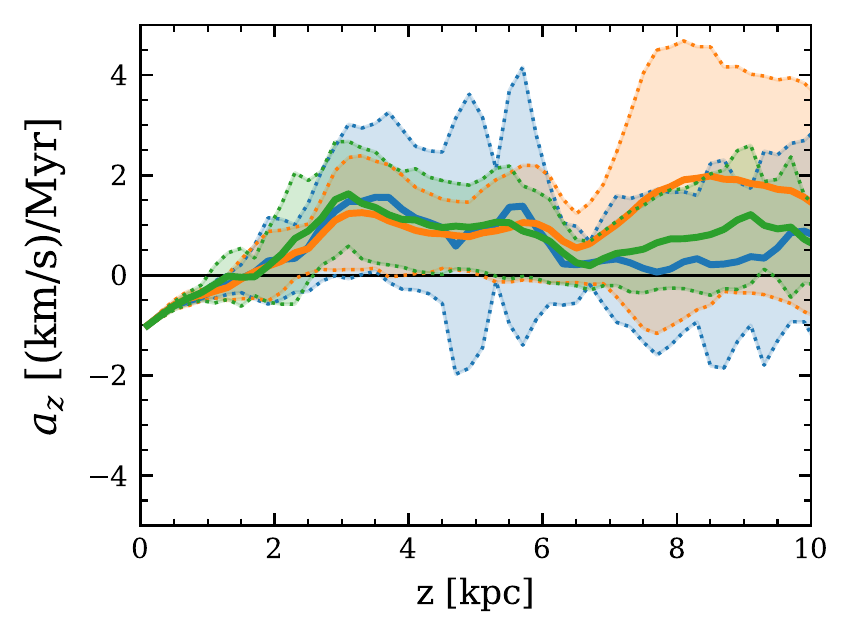}
    \caption{Vertical profile of gas acceleration in the CGM for the simulations, computed from the pressure gradient and an analytical approximation of the gravitational potential of each galaxy. The filled lines show the mean acceleration from all outputs in the final $200\Myr$, with the shaded regions showing the standard deviation.}
    \label{fig:acceleration}
\end{figure}

\begin{figure*}
    \centering
    \includegraphics[width=0.75\linewidth]{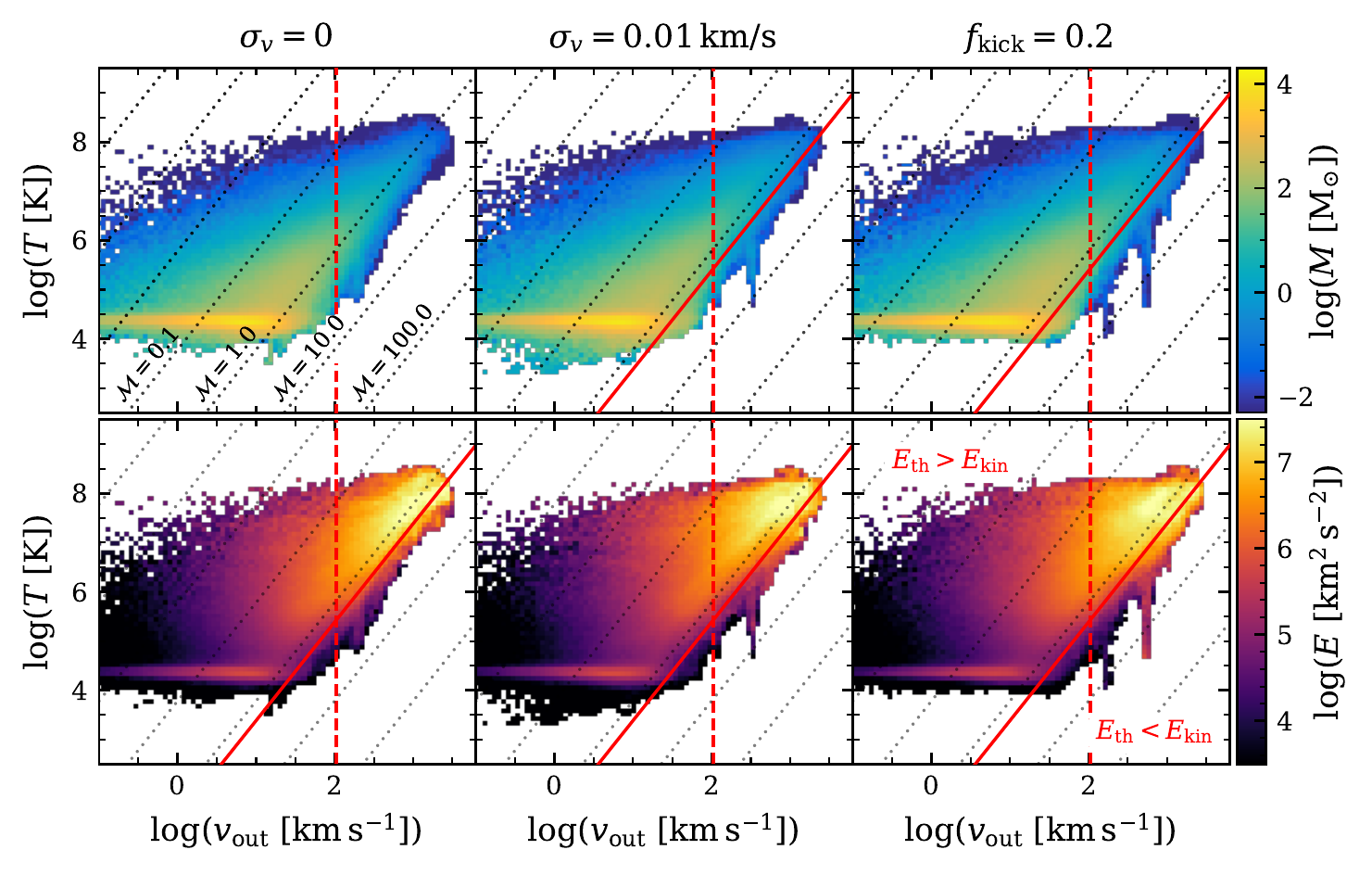}
    \includegraphics[width=0.75\linewidth,trim={0 0 0 0.6cm},clip]{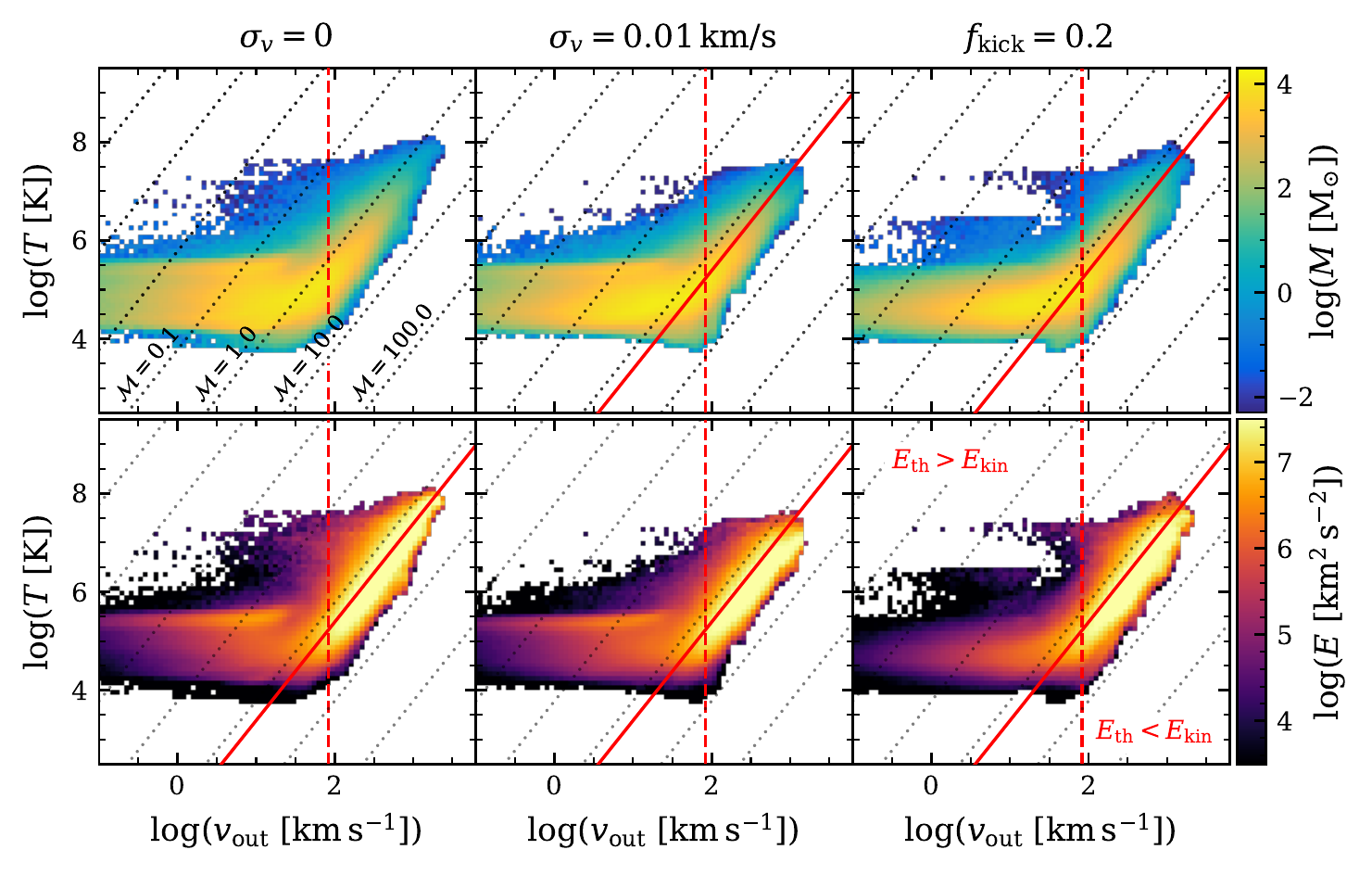}
    \caption{Temperature-velocity diagrams for outflowing gas shown for the \launching (\innercgm) interface in the top (bottom) two rows. We weigh the maps by mass or specific energy, as indicated by the color bar on the right-hand side of each row. Each panel shows the time average of the 2D-histograms for all outputs in the last $200\Myr$. Each column shows the simulation indicated by the column title. The dotted black lines draw order of magnitude Mach numbers ($\mathcal{M}=v_{\rm out}/c_{\rm s}$) calculated from the sound speed of ionized gas ($c^2=k_{\rm B}T/\mu m_{\rm H}$, with $\mu=1/2$). Solid red lines indicate where the kinetic ($0.5v^2$) and thermal ($2.5P/\rho$) energy of the outflowing gas is equivalent. The red dashed line shows the escape velocity of the dark matter halo at the location of the interface.}
    \label{fig:vT}
\end{figure*} 

\begin{figure*}
    \centering
    \includegraphics[width=0.6\linewidth]{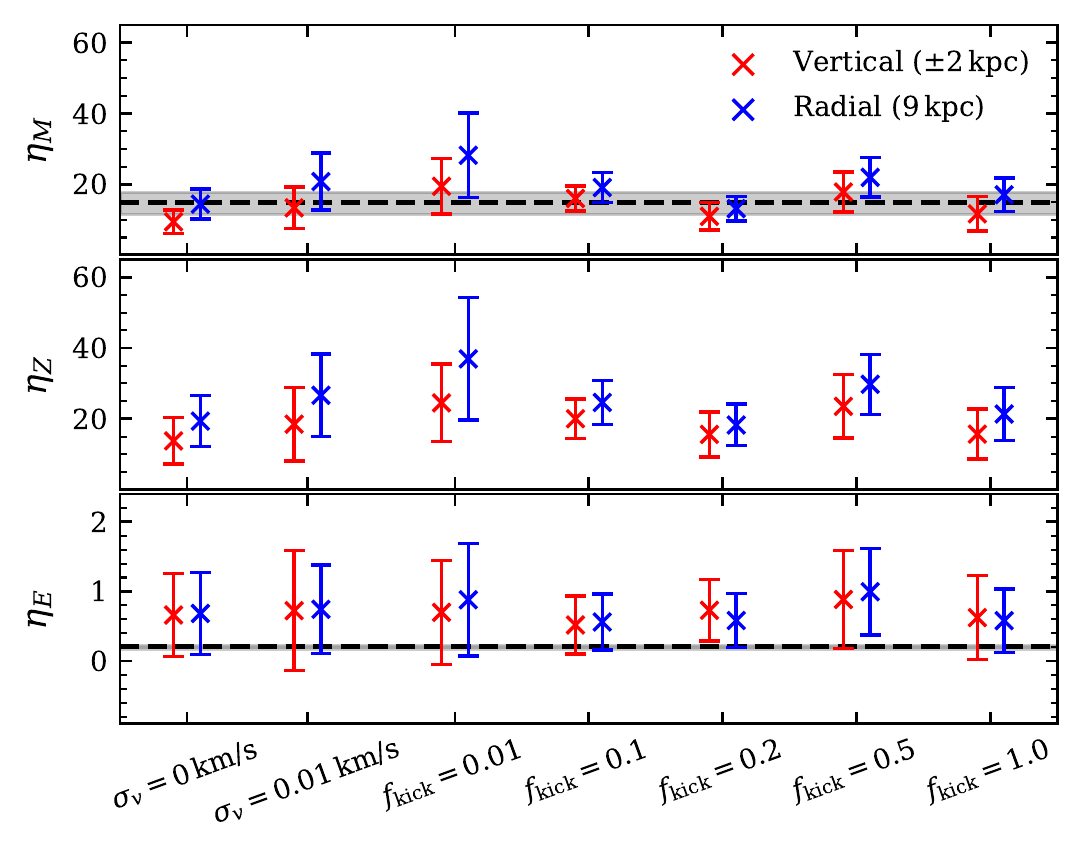}
    \caption{Mean and standard deviation of mass (top), metal (center), and energy (bottom) loading factors taken over the last $200\Myr$ of all our simulations. Red markers show the loading factors as measured at the \launching interface, while blue points show that measured at the \innercgm interface. Note that all models including some fraction of stars with the walkaway and runaway velocity distribution ($f_{\rm kick}\neq0$) also impose the $\sigma=0.01\kms$ distribution on stars that are not kicked. The black dashed line (errors indicated by gray region) shows the value from the empirical fitting function of \citet{Chisholm+2017}.}
    \label{fig:eta_allruns}
\end{figure*}

\subsection{Time-averaged wind properties}\label{sec:tave_wind}

We now turn to time-averaged properties of the outflows, considering only the final $200\Myr$ of each simulation. Note that for loading factors (see Equation~\ref{eqn:loading_factor}) we consider the fraction of the mean of the numerator and denominator separately, rather than the mean of the loading factor itself. This alleviates the problem of ill-defined loading factors when the denominator is zero.  Furthermore, we do not account for the scatter in SFR, but only consider that of the outflow. A similar approach is sometimes used in the literature when quoting loading factors resolved in time \citep[see e.g.,][]{Hislop+2022,Steinwandel+2022}

\fig{eta_prof} shows the average loading factors as a function of vertical distance (left) and radius (right). In addition to mass, metal, and energy loading factors we also include momentum loading factor\footnote{Momentum loading factor is measured by dividing $\dot{p}$ (computed from Equations \ref{eqn:qdot_z} \& \ref{eqn:qdot_r}, with $q_i=m_iv_i$) by the product of star formation rate and $1.25\times10^5/95.5\kms$. See \cite{Kim+2020b} for details on the normalization.}, as well as the energy loading factor split into thermal and kinetic energy. As previously mentioned, we see that while the vertical profile decreases with distance, the radial remains roughly constant. The decrease in vertical loading factors comes from the cylindrical slabs with a constant radius being unable to capture the full extent of the conical outflow, as these slabs are moved out. The outflow transition from thermally dominated to kinetically dominated around $z=3\kpc$ in all simulations. This is not only the result of gas cooling but also gas acceleration (seen as an increase in momentum loading). The acceleration arises due to pressure gradients existing in the halo, shown in \fig{acceleration}. We compute this by taking the gradient of the pressure profile and subtracting the gravitational force from an analytical NFW profile \citep[][the disc has negligible contribution to the potential outside $2\kpc$]{NFW1996}. We find that the acceleration becomes positive around $2\kpc$, and flattens at a value of $1\kms\Myr^{-1}$ around $3\kpc$. The flattening coincides with the transition between thermally and kinetically dominated gas energy. Provided that the acceleration can proceed far out in the halo, it can accelerate gas to $100\kms$ in $100\Myr$. As already indicated by the similarity in mass outflow rate (\fig{outflows}), there is little difference between the models. Furthermore, this affects the velocity structure of the gas between the \launching and \innercgm interfaces, described below. 

\fig{vT} shows the velocity and temperature structure of the outflows. The figure is divided into two sets of subplots, the two rows on top show the \launching interface, while the two bottom rows show the \innercgm interface. In the \launching interface, we find that outflows with temperature $T\lesssim5\times10^4\K$ dominate the mass budget (first row), while hotter outflows dominate the energy budget (second row). The majority of the mass resides in gas with velocities up to $100\kms$. Above $100\kms$, the temperature of the gas increases along a trend of roughly constant Mach number $\mathcal{M}=1.0$. This increase roughly coincides with the escape velocity of the dark matter halo. For the \launching interface, this trend only includes a small fraction of mass. At the peak of the trend, we find most of the energy, at temperatures around $\sim10^8\K$ and velocities ${\gtrsim}1000\kms$. This is in broad agreement with \citet{Kim+2020b}, who find a similar dichotomy in cold and hot gas when comparing the mass and energy budget of these different phases \citep[see also][]{Rathjen+2022}. 

When the gas reaches the \innercgm, more mass has been entrained into the fast and hot phase of the wind. We also find that the energy transitions toward more kinetic (the transition is indicated by the filled red line in \fig{vT}), likely driven by gas thermalization. The trend along a constant Mach number appears clearly, in particular in energy-weighted velocity-temperature space. As in the \launching interface, the gas is limited to subsonic velocities. 

Finally, we summarize the mass, metal, and energy loading factor for all simulations in \fig{eta_allruns}, including those presented in Appendix \ref{sec:all_runs}. The mass and energy loading factors of all our simulations are compared to the values from empirically derived fitting functions by \citet{Chisholm+2017}, shown with black dashed lines. A similar fitting function for metal loading factor is presented in \citet{Chisholm+2018}, however, our values underestimate these by 2 orders of magnitude, hence we omit including these estimates on the linear vertical axis of \fig{eta_allruns}. We do not find large differences among our simulations, but rather that all simulations have $\eta_M\sim 5-40$, $\eta_Z\sim 10-60$, and $\eta_E\sim0-2$. The largest value and scatter are found in $\sigma_v=1\kms$, and in $f_{\rm kick}=0.5$ when including walkaway and runaway stars. The minor role of runaway stars is likely a result of highly porous ISM, as well as a halo that is highly energetic. We discuss this and other factors which might affect the small role of runaway stars in \sect{kick_discussion}.


\section{Discussion}\label{sec:discussion}

The results covered have been focused on the \inferno model's ability to regulate star formation and drive galactic scale outflows via stellar feedback. For a dwarf galaxy, our model generates a strong steady outflow, with large ($>10$) mass and metal loading factors, as well as the energy loading factor close to unity (summarised in \fig{eta_allruns}). When resolved in time, we find that the loading factors display strong fluctuations (two orders of magnitude) as a result of bursty SFR. High loading factors on the dwarf mass scale are necessary to reproduce the faint end of the galaxy mass function \citep{NaabOstriker2017}, a notion that is also supported empirically \citep{Chisholm+2017,Chisholm+2018,Schroetter2019}. The outflows are more metal-rich compared to the ISM, however, not to the extent found by \citet{Chisholm+2018}. We investigate this further in Andersson et al. (in preparation), where we present the full chemical evolution model implemented in \inferno. Our outflows are highly multi-phase in nature, covering a large range of temperatures. This is crucial for the degree of ionization in the CGM \citep{Tumlinson+2017}. In stark contrast to previous results obtained for massive disc galaxies \citep[][]{Andersson+2020}, we find that the natal velocity distribution of the stars plays a minor role in setting the loading factor on dwarf scales. In the following sections, we discuss this in more detail. 

\subsection{Comparisons with contemporary feedback models}\label{sec:discuss_comparison}

Our simulation setup of a dwarf galaxy in an isolated environment allows us to reach parsec scale resolution, which is comparable to works by e.g. \citet{Hu2019,Emerick+2020,Smith+2021,Steinwandel+2022}. While the initial gas mass of these models varies (particularly in \citeauthor{Emerick+2020}, who simulated an ultra-faint dwarf), the mass and metal loading factor are in broad agreement. However, the energy loading factor of our simulations is around unity, while the aforementioned works routinely find values around $0.1$. As discussed below, this discrepancy by an order of magnitude could provide insight into differences in feedback models and numerical treatment. The energy supplied to the halo affects the re-accretion of material, dividing feedback into {\it preventive} (inhibiting gas inflow) and {\it ejective} (expelling gas) feedback \citep{Dave+2012}. For a thorough literature comparison, we refer to \citet{Li&Bryan2020}. 

Of particular interest is the work by \citeauthor{Smith+2021} (\citeyear{Smith+2021}, see also \citealt{Smith2021}). \citeauthor{Smith+2021} investigated a suite of simulations with similar ICs and numerical resolution, but with a different hydrodynamics solver, star formation recipe, and for a range of different feedback sources \citep[see also][]{Hu2019}. The SN-PE model in \citeauthor{Smith+2021} is the most comparable in terms of included feedback processes (although we include SNIa and stellar winds, which may be of importance, see \sect{SNIa}), with which we find slightly lower mass loading (factor 2), but a significantly higher energy loading (factor 10). Note that when \citeauthor{Smith+2021} introduces radiation feedback, the energy loading decreases significantly, thereby increasing the discrepancy with our model. The origin of the discrepancy between our model and that of \citeauthor{Smith+2021} is not clear; however, \citet{Hu+2022} found notable differences between \ramses and \arepo (used in \citealt{Smith+2021}). \citeauthor{Hu+2022} attributed these differences primarily to star formation and its effects on clustered SNe. In our case, a higher star formation efficiency can account for this due to stronger clustering of SNe \citep{Hu+2022}. Note that \cite{Smith+2021} explored different sub-grid prescriptions for star formation (e.g., changing the star formation efficiency) and concluded that their results were insensitive to such changes, although this is not clear for our model. Concerning energy injection, \ramses updates the energy in a fixed volume (set by the refinement level), while (quasi-) Lagrangian codes \citep[e.g.,][the latter using the moving-mesh method of \arepo]{Hu2019,Smith+2021} often update energy in regions of fixed mass. This difference may affect the injection feedback, e.g., adiabatic cooling processes or spatial clustering of injection events. Likely, the numerical method plays other roles as well, a factor which has been discussed extensively in literature \citep[see, e.g.,][]{AGORA2014,AGORA2016,AGORA2020,Hu+2022}. Further efforts to understand the differences between codes (particularly for star-by-star models) are likely necessary to reach a consensus.

\subsection{The weak impact of natal stellar kinematics on stellar feedback}\label{sec:kick_discussion}

\subsubsection{The role of disc structure and the ISM}
As previously mentioned, the density structure of the ISM likely plays a role in how relevant runaway stars are. In \citet{Andersson+2020}, the inclusion of runaway stars resulted in a supply of CCSNe progenitors into large under-dense regions which enabled SNe to more efficiently drive outflows \citep{Ohlin+2019}. This could also explain the disagreement found by \citet{Kim+2020b}, which does not capture the low-density regions imparted by spiral arm shearing and the full geometrical extent of the galaxy \citep[see also][]{Martizzi+2016}. In the simulations presented here, the low SFR implies that the number of runaway stars is low, hence, although low-density regions develop, they are unlikely to receive a significant number of runaway stars before dissolving. We confirmed this through a visual inspection of our simulations. 

Furthermore, the shallow potential of the disc in the dwarf galaxy implies a thick gas disc (initial scale height $h_{\rm disc}=0.7\kpc$). As such, stars need to travel a long distance to reach dramatically different environments, even when traveling vertically (unless reaching far out in the CGM, see \sect{CGMrunaways}). In contrast, runaway stars in more massive disc galaxies (e.g. as in \citealt{Andersson+2020} with a gas disc scale height of $h_{\rm disc}=0.34\kpc$) have a shorter travel distance to environments with as dramatic differences (e.g. in gas density).

The weak impact of runaway stars could be connected to the star formation threshold ($500\cc$). Such a high density implies that star formation depletes the local gas reservoir on a timescale $\tau_{\rm dep}=\rho_g/\dot{\rho}_{\rm sf}=t_{\rm ff}/\epsilon_{\rm ff}\simeq20\Myr$, which is similar to the timescale for SNe. Varying the star formation threshold and $\epsilon_{\rm ff}$ affects the clustering of SNe and to some extent the outflow properties \citep{Smith+2021}. If gas is depleted fast, massive stars in clusters explode as SNe in low-density gas, leading to efficient heating of the ISM \citep[e.g.,][]{Agertz&Kravtsov2015}. Differences in star formation recipes (as well as numerical resolution, see discussion in \citealt{Kim+2020a}) could be the cause for the discrepancy in the results of e.g., this work, \citet{Kim&Ostriker2018}, \citet{Andersson+2020}, and \citet{Steinwandel+2022}. This would explain the lack of consensus regarding the impact of runaway stars. However, the full explanation is likely more intricate, with many factors playing a role, e.g. overall structure of the ISM (which determines escape channels for SNe energy, see e.g., \citealt{Hayward&Hopkins2017,Ohlin+2019}), or pre-SNe feedback (which can counteract gas collapse, see e.g., \citealt{Smith+2021}).

\subsubsection{The role of type Ia supernovae}\label{sec:SNIa}

\begin{figure}
    \centering
    \includegraphics[width=\linewidth]{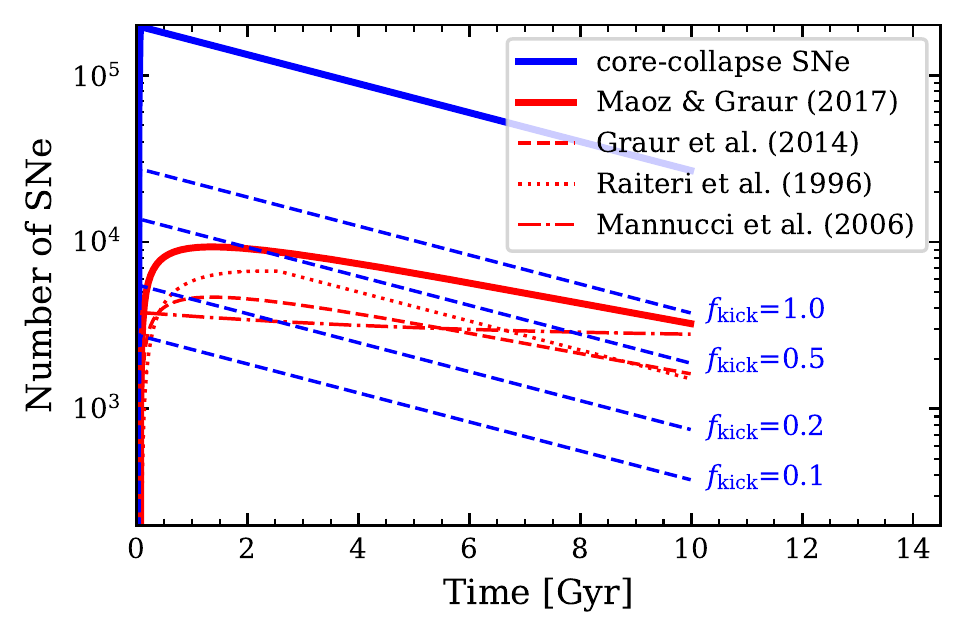}
    \caption{Number of CCSNe (blue) and SNeIa (red) as a function of time calculated for an exponential star formation rate. The dashed blue lines show the number of CCSNe related to runaway stars, assuming different cluster escape fractions $f_{\rm kick}$ ($=1$ implies $14\%$ runaway stars), labeled in the figure. The different red lines show SNeIa rates for different models widely used in the literature. The filled red line is the one used in our simulation.}
    \label{fig:IaRA}
\end{figure}

\begin{figure}
    \centering
    \includegraphics[width=\linewidth]{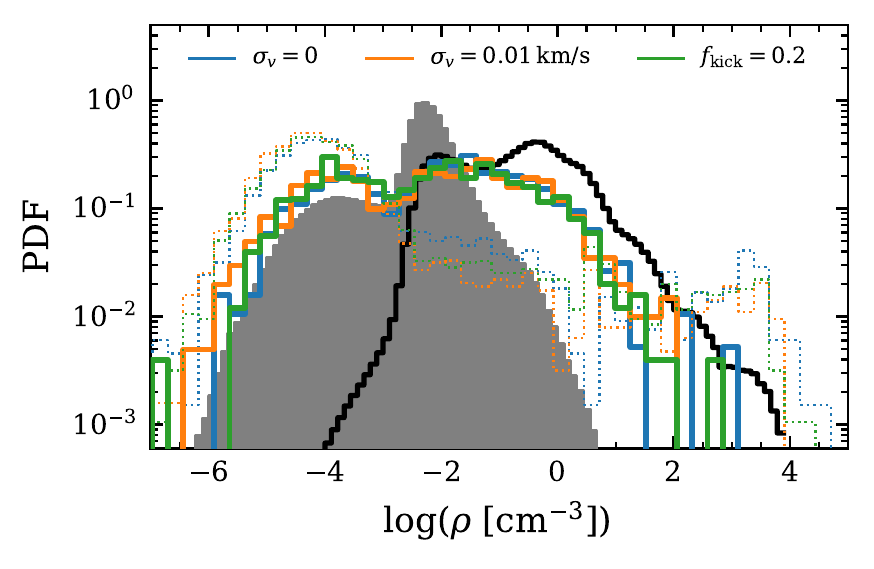}
    \caption{The probability distribution function of density, with colored lines showing the densities where SNeIa explodes in our simulations. As in Figure 6, the filled grey histogram (thick black line) shows the volume (mass) weighted density distribution for the $f_{\rm kick}=0.2$ model. For comparison, we included the distribution of CCSNe with thin dotted lines. Note that the densities are sampled at different cadences; Ia densities are recorded at the time of the explosion, while the mass- and volume-weighted densities are computed at the $10\Myr$ output rate.}
    \label{fig:Ia_dens}
\end{figure}

A key aspect of the supposed effect that massive runaway stars have on stellar feedback is that they explode far away from where they were formed. This leads to more randomly distributed SNe sites, in contrast to SNe only around star-forming gas \citep[see, e.g.,][]{Li+2015,Li+2017}. To a large extent, this is also the case for delayed SNe, e.g. SNeIa with rates which are a few tens of per cent of the CCSNe \citep{Tammann+1994}. Exploring the role of these SNe in the context of dwarf galaxies warrants follow-up work, but we can speculate on their effect since our model implements a method to include these objects. Note that at late times in our simulations (final $200\Myr$), SNeIa makes up $\sim20\%$ of the total SNe population. With only a small per cent of stars being runaways, SNeIa may in fact be the main contributor to randomly located stellar explosions. To exemplify this for an extended star formation history, we show the number of CCSNe and SNeIa in \fig{IaRA}, assuming a simple toy model with exponential ${\rm SFR}=\exp(-t/5\Gyr)\msunyr$. Also shown is the number CCSNe associated with runaway stars for different values of $f_{\rm kick}$, as well as several different models for the SNIa rate. In our simulations, we adopt the model by \citet{Maoz&Graur2017}. The second model shown is from \citet{Graur+2014}, which is the same as \citet{Maoz&Graur2017} but normalized to fit data from galaxy clusters (half the rate of field galaxies). The model by \citet{Raiteri+1996} has been widely used in early galaxy models \citep[see e.g.,][]{Greggio&Renzini1983,Matteucci&Greggio1986,Agertz+2013}. We also show the model derived by \citet{Mannucci+2006} used in the \fire model for galaxy simulations \citep{Hopkins+2014,Hopkins+2018,Gandhi+2022}.

The SNIa rate builds up in the first few $\Gyr$, and is comparable to the number of SNe associated with runaway stars even in models assuming a high fraction of kicked stars ($f_{\rm kick}\geq0.5$). To our best knowledge, the role that this build-up of SNeIa has on non-cosmological simulations is not well explored in the literature.

We show the gas densities where SNeIa explodes in \fig{Ia_dens}. Interestingly, we find that the density distribution of SNeIa explosions is a combination of the mass- and volume-weighted density PDF. If distributed homogeneously, one expects this distribution to follow the volume-weighted one. Nonetheless, we find the Ia explosions to extend toward higher densities. Furthermore, we find no correlation between the age of the star particle when the Ia occurs and the explosion density, implying that even early Ia is no longer associated with any particular density. 

If SNeIa affects the role of runaway stars, this might explain some of the discrepancies between our results and those of e.g. \citet{Smith+2021,Steinwandel+2022}.

\subsubsection{CGM and out-of-disc runaway stars}\label{sec:CGMrunaways}

\citet{Steinwandel+2022} finds that \emph{out-of-disc} runaway stars can supply a significant amount of energy, thereby increasing the energy loading in simulations that include runaway stars. We do not find this to be the case in our simulations, despite us studying a galaxy of similar mass, and we see runaway stars escaping into the halo (see \fig{TSNe}). It is likely that the high energy loading of all our simulations creates an environment around the galaxy that makes additional thermal energy dumps negligible. Indeed, in the case of $f_{\rm kick}=0.2$, the CGM inside the \innercgm interface contains $\sim400\times10^{51}\erg$ of thermal energy throughout its later evolution, i.e. hundred times larger than what any single SNe would provide. Compared to this highly energy-loaded halo, \emph{out-of-disc} runaway stars only supply a small amount of energy (we find only $\sim40$ SNe $200\pc$ above the disc in the final $200\Myr$ of our $f_{\rm kick}=0.2$ simulation).

Another aspect of such a high thermal energy content is that this establishes a negative pressure gradient. As shown in \sect{tave_wind}, this results in an outward acceleration of significant amounts of gas ejected from the galaxy. Furthermore, it is likely that material is accelerated to high velocities by SNe blast waves ($1000\kms$) that break out from the disc. The details of this will be explored in future work. 

\subsection{Limitations of \inferno}
This article presents the first iteration of the \inferno model. Therefore the model has several remaining limitations. This section highlights the most prominent of these limitations.

With the ability to reduce the clustering of star formation \citep[see, e.g.,][]{Hislop+2022}, radiation feedback typically limits loading factors, particularly in energy \citep{Smith+2021}. Furthermore, \citet{Agertz+2020} showed that radiative feedback strongly affects the formation of ultra-faint dwarf galaxies (by significantly suppressing star formation, radiation leads to an overall calmer evolution). \inferno does not yet include radiation feedback, although stellar winds play a similar (but weaker) role (Andersson et al., in preparation). Radiative hydrodynamics are already implemented in \ramsesrt \citep[][]{Rosdahl+2013,Rosdahl+2015}. This model is currently being adapted for \inferno and will account for radiation feedback using stellar spectra from individual stars to employ star-by-star radiation feedback. How this affects runaway stars remains unclear.

The star formation recipe employed by \inferno relies on IMF sampling from discrete quanta of stellar material $M_{\rm sf}$. To ensure accurate sampling, $M_{\rm sf}$ is constrained by $\gtrsim500\Msun$ \citep[see, e.g.,][]{Smith2021}. This mass constraint implies that an increasing resolution forces higher gas density to allow star formation. How this affects the formation of stars in our simulations is not clear. Solutions to this problem are to either abandon predefined mass bins for IMF sampling and immediately sample stars from the IMF \citep[see, e.g.,][]{Lahen+2019} or to introduce sink particles to model the star formation process at scales smaller than the resolution elements \citep[see, e.g.,][]{Bate+1995,Klassen+2016,Gatto+2017}. An advantage of introducing sink particles is that this allows a more straightforward way to model time-resolved star formation \citep[e.g., the delayed formation of massive stars][]{Haugbolle+2018} and time-dependent natal velocity kicks \citep[see, e.g.,][]{OhKroupa2016}.

Because collisional gravitational dynamics are far from resolved in our simulations, our model is not predictive concerning the physics of binary stars. The sub-grid model for runaway stars and SNe type Ia only depends on binary objects implicitly. Furthermore, stellar multiples are necessary to explain exotic astrophysical objects, such as stripped envelop stars, SNe kicks, and stellar mass transfer \citep[see, e.g.,][]{Hurley+2002,Izzard+2006}. These aspects affect, e.g., stellar feedback, chemical enrichment, and stellar kinematics, and are therefore of interest to investigate further in \inferno. In the context of stellar multiplicity, Blaauw kicks \citep[][]{Blaauw1961} are particularly interesting for our model since these are a source of runaway stars. Blaauw kicks are triggered by immediate mass loss in a binary system when the companion star undergoes SNe (note that this implies a time delay before the velocity kick). Our model does not include this effect because the velocity distribution applied only accounts for the first $3\Myr$ star cluster evolution \citep{OhKroupa2016}. Be mindful that our model adds all velocity kicks at the birth of each stellar population. Introducing binary stars via a parametrized method could allow us to explore these aforementioned physical processes without the necessity of costly collisional dynamics \citep[see, e.g.,][]{Eldridge+2011,Kim&Ostriker2017}.

Finally, \inferno remains limited by an equilibrium cooling physics \citep[see, e.g.,][for details on the effects of non-equilibrium chemistry]{Katz+2022} and lacks several physical mechanisms known to affect galaxy evolution \citep[e.g., magnetic fields and cosmic rays, see][]{NaabOstriker2017}. For example, cosmic rays can generate a pressure gradient that drives primarily cold and warm gas into the outflows \citep[see, e.g.,][]{Rathjen+2022}, which would appear as an additional phase at a lower temperature in \fig{vT}. Future efforts toward making \inferno a more detailed model for galaxy simulations will focus on these aspects.


\section{Summary \& Conclusions}\label{sec:conclusions}

We present a new galaxy physics model called \inferno, introducing a star-by-star treatment for the injection of momentum, energy, and chemically enriched material, each with timing, locality, and amount calculated based on the properties of individual stars. We employ \inferno to simulate the evolution of a dwarf galaxy to study how stellar feedback drives outflows. Our results focus on the mass, metal, and energy loading factors, as well as the properties of galactic winds.

We draw the following concluding remarks from our study:
\begin{enumerate}
    \item Our stellar feedback model causes a lowering of star formation by roughly two orders of magnitude while driving strong gas outflows. A galactic wind is established close to the disc (around $\pm2\kpc$ from the disc mid-plane) and moves material through the CGM. We recover mass and metal loading factors on the order of $10-100$, as required to match the faint end of the galaxy mass function \citep{NaabOstriker2017}. Furthermore, the wind is heavily energy-loaded, with an energy loading factor close to unity.
    \item The galactic winds display a clear dichotomy in the mass and energy outflow, with mass primarily carried by cold gas ($T\simeq10^5\K$) at velocities $v<100\kms$, while energy is carried in a hot ($T>10^7\K$), fast ($v>100\kms$) wind. The energy evolves with distance from the galaxy, transitioning from thermally dominated to kinetically dominated a few $\kpc$ above the disc plane. Our model generates a highly energetic CGM where outflows are limited to the subsonic regime, with high-velocity gas ($v>100\kms$) following a trend of roughly constant Mach number $\mathcal{M}\sim0.1$ in the velocity-temperature space. 
    \item We find no strong effects imposed by the different natal velocity distribution applied to newly formed stars. While we include runaway stars in our model, we find a surprising insensitivity to their presence, in stark contrast to more massive galaxies where runaway stars play a significant role in setting the outflows \citep{Andersson+2020}. Not only is this the case for outflows ejected by dwarfs, but we find similar SFRs, gas multi-phase structures, and SNe explosion densities, regardless of what natal stellar velocity distribution we apply.  
\end{enumerate}

The precise role played by runaway stars for galaxy evolution is not yet established, with varying conclusions in the literature \citep[][]{Ceverino&Klypin2009,Kimm&Cen2014,Andersson+2020,Kim+2020b,Steinwandel+2022}. At this stage, the literature covers a wide range of galaxy masses, which are simulated with a multitude of different models. This work is the first in a series that will employ \inferno, with the aim of exploring runaway stars, as well as galaxy evolution physics in general. 

\section*{Acknowledgements}
We thank the referee Tiago Costa for providing constructive comments which helped improve the quality of this work. EA thanks Ulrich P. Steinwandel and Martin P. Rey for insightful conversations and helpful comments.
EA, OA, and FR acknowledge financial support from the Knut and Alice Wallenberg Foundation and the Swedish Research Council (grant 2019-04659). EA acknowledges financial support from the Royal Physiographic Society of Lund and from NSF grant AST18-15461. EA acknowledges computer resources from Swedish National Infrastructure for Computing (projects SNIC 2021/5-111, SNIC 2021/6-87, and SNIC 2021/6-85) and Large Unified Modern Infrastructure (LUMI pilot phase).
\section*{Data Availability}
The data underlying this article will be shared on reasonable request to the corresponding author.



\bibliographystyle{mnras}
\bibliography{ref}



\appendix
\section{All simulations}\label{sec:all_runs}
\begin{figure}
    \centering
    \includegraphics[width=1.0\linewidth]{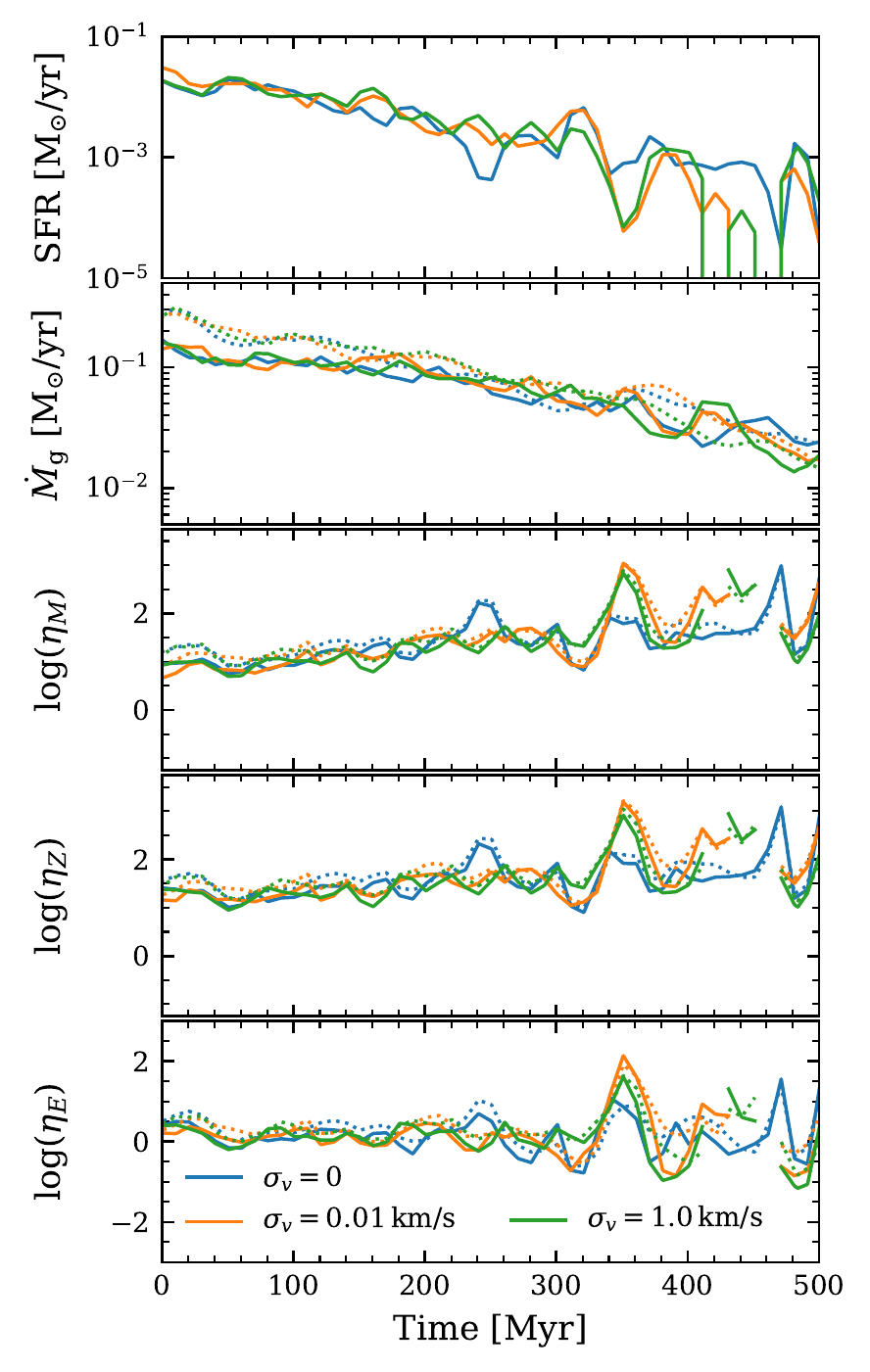}
    \caption{The top two rows show SFR and mass outflow rate, while the remaining three rows show mass, metal, and energy loading from top to bottom, all as a function of time for the \stir simulations. Rates are computed in $10\Myr$ time-bins, with filled lines showing the \launching interface and dotted lines showing the \innercgm interface. Different values of $\sigma_v$ are denoted in the legend of the bottom plot.}
    \label{fig:rates_eta_stir}
\end{figure}

\begin{figure}
    \centering
    \includegraphics[width=\linewidth]{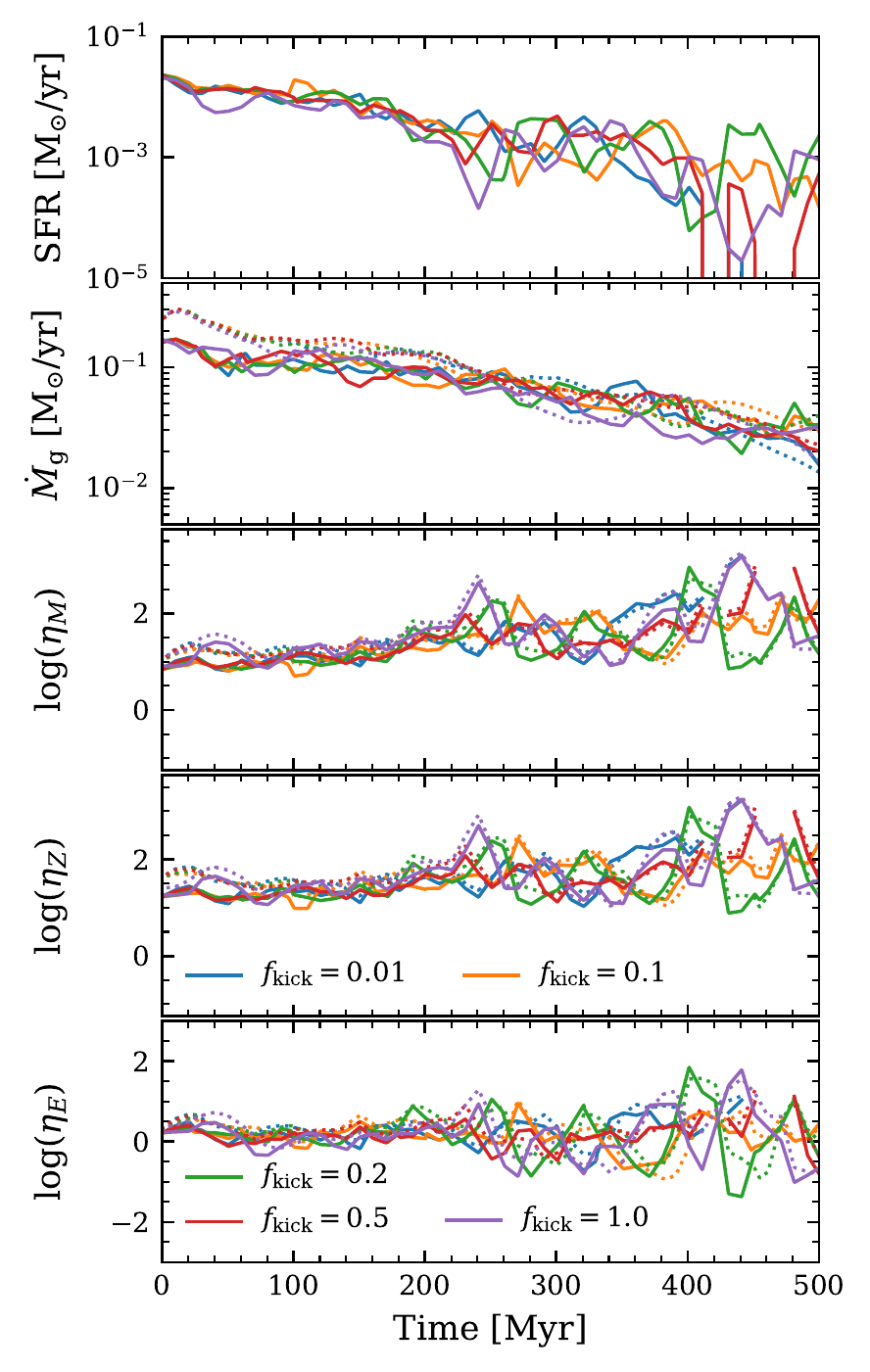}
    \caption{Same as for Figure A1, but for the \kick models. Different values of $f_{\rm kick}$ are denoted in the legends of the bottom two plots.}
    \label{fig:rates_eta_kick}
\end{figure}

Here we present the full suite of simulations aimed at extensively exploring how the natal velocity distribution of individual stars affects our feedback model. The \stir model is parameterized by $\sigma_v$, for which we tested values $0$, $0.01\kms$, and $1\kms$. This model intends to allow co-natal stars to have diverging trajectories arising from small perturbations in the gravitational potential. That the stars do not do so without \stir is a numerical effect of the collisionless particle-mesh gravity solver, and thus a small value for $\sigma_v$ is preferred. Nonetheless, our results do not change drastically between the values we tested, as shown in Figure \ref{fig:rates_eta_stir}.

The \kick model implements walkaway and runaway stars following the velocity distribution of stars escaping clusters through dynamical interactions \citep{OhKroupa2016}. Because both the fraction of stars born in clusters and the fraction of stars which escape clusters is not well known, we parameterize this with a kick fraction $f_{\rm kick}$ and apply it only to massive stars ($>8\Msun$). We tested values $0.01$, $0.1$, $0.2$, $0.5$ and $1.0$. Note that for $f_{\rm kick}=1.0$ the fraction of massive runaway stars is $14\%$. We find little to no effect from runaway stars for all values, as shown in \ref{fig:rates_eta_kick}.

\begin{figure*}
    \centering
    \includegraphics[trim={0 0.8cm 0 0},clip,width=\linewidth]{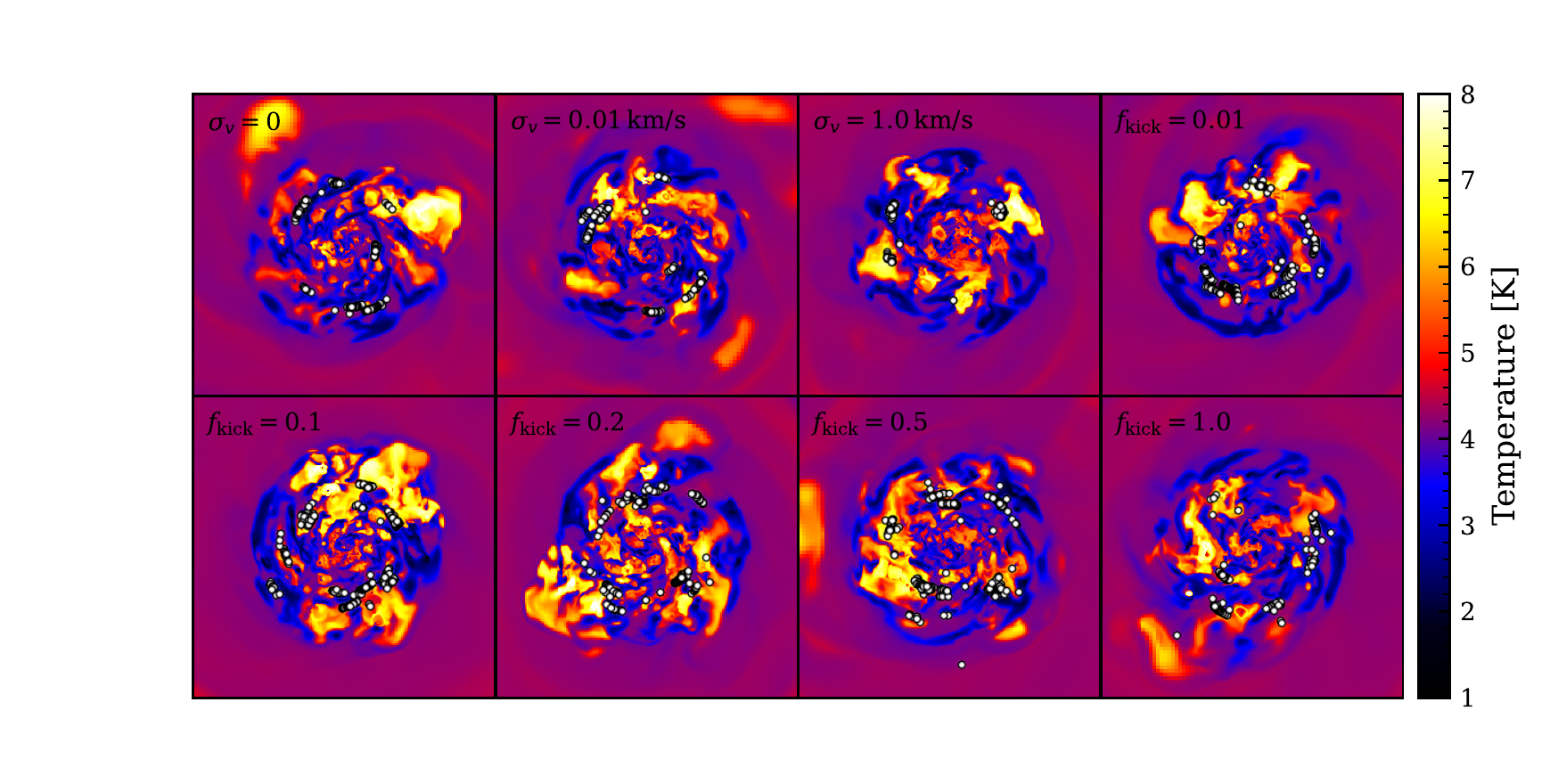}
    \includegraphics[trim={0 0 0 0.8cm},clip,width=\linewidth]{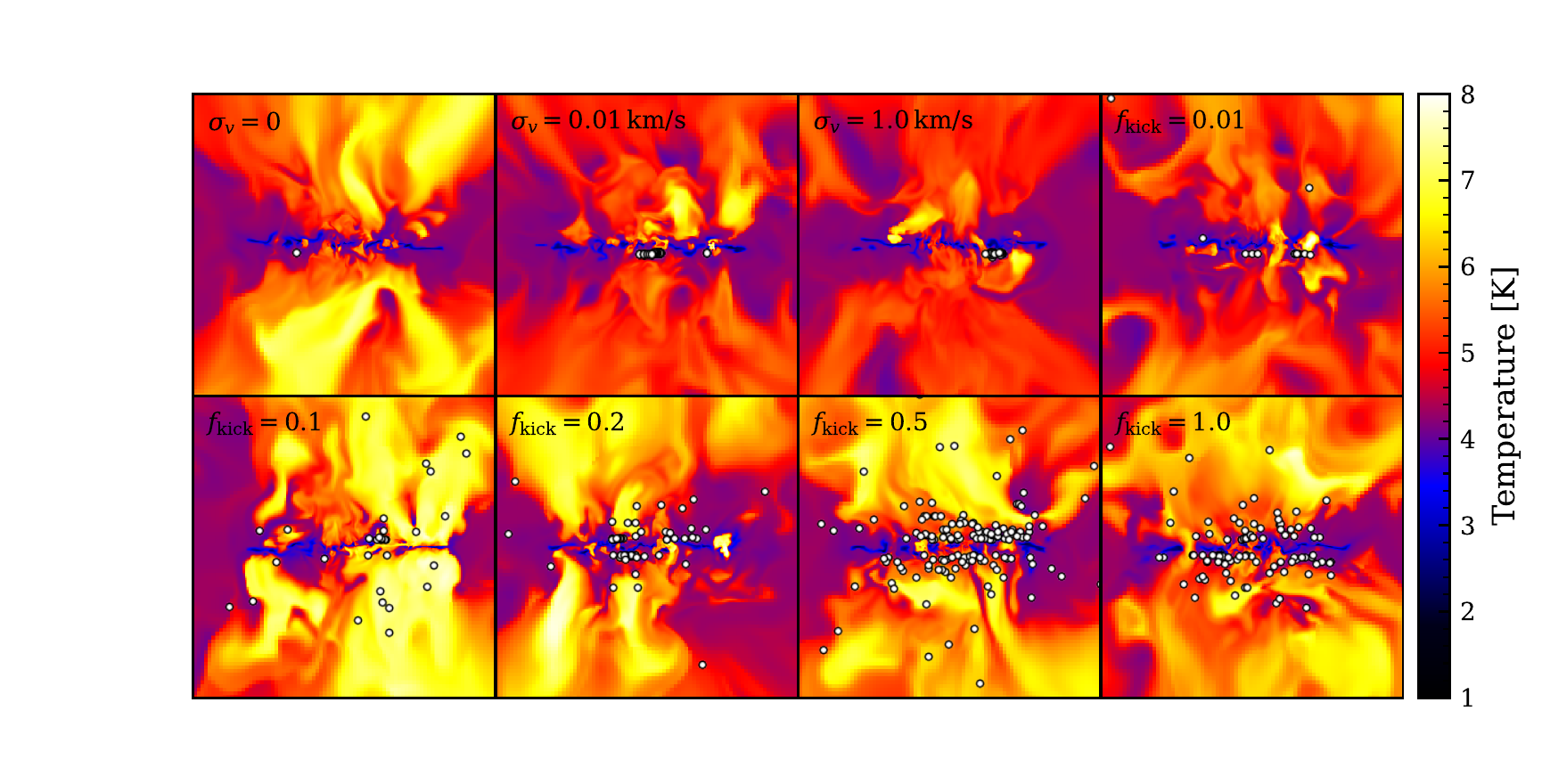}
    \caption{Gas temperature in slices through the centre of our simulation boxes showing the face-on view in the top panels and edge-on view in the bottom panels. All plots shown are for $t=400\Myr$, and the panels have a width of $8\kpc$. Each panel shows a different simulation denoted by the label in the upper left corner. The location of recent ($<5\Myr$ in the disc; $<50\Myr$ outside the disc) SNe are shown in white points.}
    \label{fig:TSNe}
\end{figure*}

Finally, in Figure \ref{fig:TSNe} we show the locations of recent CCSNe for all models with the face-on view in the upper plot and edge-on view in the lower plot. This is shown on top of the temperature maps of each simulation. Notably, we see how the number of \emph{out-of-disc} SNe increases, as we increase $f_{\rm kick}$.

\section{Resolution tests}\label{sec:convergence_test}
In this appendix, we address how sensitive our results are to the resolution of our simulations. Selecting the $f_{\rm kick}=0.2$ model, we re-simulate the evolution of the galaxy at lower resolution to see whether significant differences appear in the outflow rates and loading factors. The resolution of the simulations can be affected either by limiting the maximum level of refinement or by changing the mass criterion for when refinement is triggered. To investigate the sensitivity to both these refinement criteria, we simulated one galaxy with a maximum refinement level at 13 levels (i.e., at three levels lower, resulting in a max spatial resolution of $12\pc$) and one where the refinement mass increased by a factor 10 (i.e., a refinement mass of $8\times1000\Msun$). Note that decreasing the numerical resolution affects the highest densities (due to the smoothing of gravitational forces). Since the star formation recipe operates at the resolution limit, the star formation threshold must be adapted to provide a similar star formation history. There is no straightforward method for this. However, through tests, we find that changing the star formation threshold to $10\cc$ in the case when the spatial resolution is limited to $12\pc$ and $50\cc$ for a refinement mass of $8\times1000\Msun$ provides star formation rates that are similar to the simulations at the original simulations (we also present results from simulations where the star formation threshold remained unchanged).

\begin{figure}
    \centering
    \includegraphics[width=\linewidth]{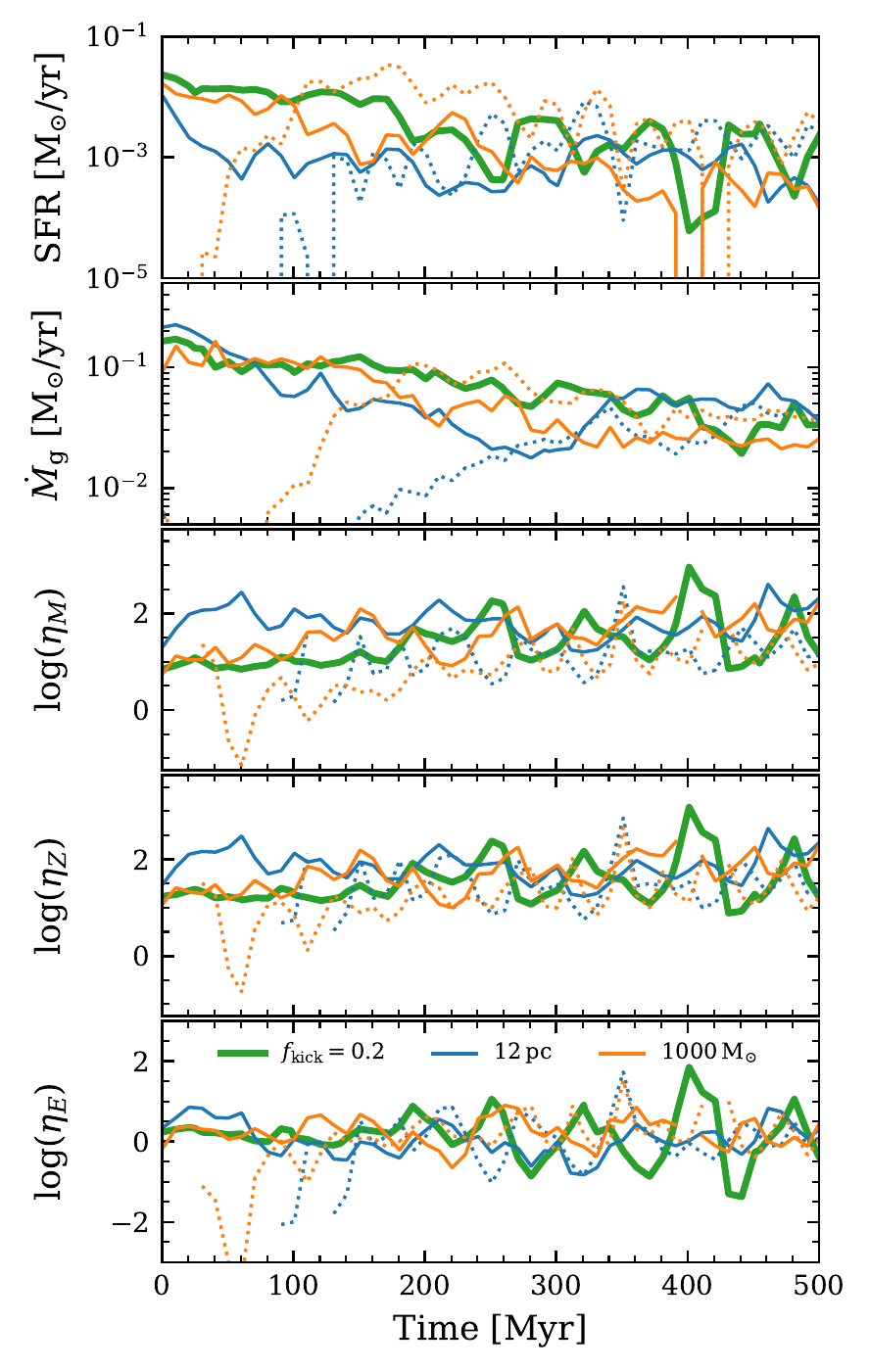}
    \caption{Star formation rate, outflow rate, mass, metal, and energy loading shown from top to bottom as a function of time for the $f_{\rm kick}=0.2$ simulation at a different resolution. The green, thick line shows the results presented in the main body of the article (spatial resolution of $\sim1.5\pc$, and mass resolution of $100\Msun$). The thin lines show simulations at lower resolution ($12\pc$ in blue and $1000\Msun$ in orange). The star formation density threshold is $10\cc$ and $50\cc$ in $12\pc$ and $1000\Msun$, respectively (see main text for details). The dotted lines show simulations with a star formation density threshold identical to the original simulations ($500\cc$).}
    \label{fig:convergence}
\end{figure}

In Figure~\ref{fig:convergence}, we show the star formation rate, outflow rate, and loading factors as a function of time for the simulations at different resolutions. Note that at early times, not adapting the star formation threshold results in a slower build-up of the star formation rate and, consequently, the outflow rate. Nonetheless, the loading factors are less affected by this, indicating that the energy and momentum budget of the feedback model remains similar. Note that in the case of $12\pc$ ($8\times1000\Msun$), roughly $10\%$ ($15\%$) of SNe have unresolved Sedov-Taylor evolution). At later times, the wind displays similar rates in all simulations. 

While the changes to the resolution in the most highly resolved parts of the galaxy (e.g., the ISM) do not drastically affect our results, this is not necessarily the case for the CGM. As shown in \citet{Rey+2023}, increasing the resolution outside the galaxy (which is inherently low due to the nature of the AMR) can significantly boost the outflow rates. How strongly this affects the results presented here and in all other work focusing on outflows remains to be seen.

\bsp	
\label{lastpage}
\end{document}